\documentclass[aps,pre,twocolumn,groupedaddress]{revtex4-2}
\usepackage{amsmath}
\usepackage{graphicx}
\usepackage{natbib}
\usepackage{bm}
\usepackage{color}

\newcommand{\bec}[1]{\mbox{\boldmath $ #1$}}

{}
\newcommand{\meanUU}{\overline{\bm{U}}}

{}
{}

\newcommand{\meanT}{\overline{T}}

\newcommand{\meanU}{\overline{U}}

\begin{document}

\title{The energy- and flux budget theory for surface layers in atmospheric convective
and stably stratified turbulence
}

\author{I.~Rogachevskii$^{1,2}$}
\email{gary@bgu.ac.il}
\homepage{http://www.bgu.ac.il/~gary}
\author{N. Kleeorin$^{1,2}$}
\author{S. Zilitinkevich$^{3,4}$}

\bigskip
\affiliation{
$^1$Department of Mechanical Engineering, Ben-Gurion University of the Negev, Beer-Sheva
 8410530, Israel
 \\
$^2$Nordita, Stockholm University and KTH Royal Institute of Technology, 10691 Stockholm, Sweden
 \\
$^3$Institute for Atmospheric and Earth System Research (INAR), University of Helsinki, 00014 Helsinki, Finland
 \\
$^4$Finnish Meteorological Institute, 00101 Helsinki, Finland
}

\date{\today}
\begin{abstract}
The energy- and flux budget (EFB)  turbulence closure theory for the atmospheric surface layers in convective
and stably stratified turbulence has been developed using budget equations for turbulent energies and fluxes in the Boussinesq approximation.
In the lower part of the surface layer  in the atmospheric convective boundary layer (CBL),
the rate of turbulence production of the turbulent kinetic energy (TKE) caused by the mean-flow surface shear and the shear of self-organised
coherent structures is much larger than that caused by the buoyancy, which results in three-dimensional turbulence of very complex nature.
In the upper part of the surface layer, the rate of turbulence production of TKE due to the shear is much smaller
than that caused by the buoyancy, which causes unusual strongly anisotropic buoyancy-driven turbulence.
Considering the applications of the obtained results to the atmospheric convective and stably stratified boundary-layer turbulence,
the theoretical relationships potentially useful in modelling applications have been derived.
In particular, the developed theory for the surface layers  in turbulent convection and stably stratified turbulence
allows us to determine the vertical profiles for all turbulent characteristics,
including TKE, the intensity of  turbulent potential temperature fluctuations, the vertical turbulent fluxes
of momentum and buoyancy (proportional to potential temperature), the integral turbulence scale,
the turbulence anisotropy, the turbulent Prandtl number and the flux Richardson number.
\end{abstract}


\maketitle

\section{Introduction}
\label{sec:1}

In spite of turbulent transport have been studied theoretically, in laboratory and field
experiments and numerical simulations
during a century \citep{MY71,MY75,MC90,F95,P2000,LE08,D13,RI21},
some crucial questions remain.
This is particularly true in applications such as geophysics and astrophysics,
where the governing parameter values are too
large to be modelled either experimentally or numerically.
The classical Kolmogorov’s theory has been formulated for a
neutrally stratified homogeneous and isotropic turbulence
\citep{K41,K41b,K42,K41a}.
This turbulence is different from convective  and stably stratified turbulence.

Modern understanding of atmospheric convective turbulence is based on
the following \citep{ZRK21}:
\begin{itemize}
\item{
buoyancy produces chaotic vertical plumes,
which are different from small-scale turbulent eddies;
}
\item{
the small-scale  turbulent eddies which are produced by the mean-flow shear and the shear of self-organised
coherent structures,
are unstable and break down in  smaller unstable eddies, thus causing the direct
cascade of the turbulent kinetic energy;
}
\item{
merging of small plumes into larger plumes results in
an inverse energy cascade towards their conversion into
the self-organized large-scale coherent structures;
}
\item{
 in convective turbulence,  there are countergradient and nongradient
turbulent transports.
The gradient transport of momentum,
energy, and matter implies that the turbulent flux of any quantity is determined by
the product of the mean gradient of the transferred quantity and
the turbulent transport coefficient.
The nongradient turbulent transports means that the turbulent flux
is not determined by the mean gradient of the transferred quantity.
}
\end{itemize}
This concept is based on various experimental, numerical and theoretical studies
\citep{Z71,Z73,Z91,ZHE06,WC71,KY90,EKRZ02,EKRZ06,EGKR06,GD13,BK15,SCB17,SA18,SA20}.
The atmospheric turbulent convective boundary layer (CBL) consists in three basic parts:
\begin{itemize}
\item{
Surface layer strongly unstably stratified and dominated by small-scale turbulence of
very complex nature including usual 3-D turbulence, generated by mean-flow surface
shear and structural shears (the lower part of the surface layer), and unusual strongly
anisotropic buoyancy-driven turbulence (the upper part of the surface layer);
}
\item{
CBL core dominated by the structural energy-, momentum- and mass-transport, with only
minor contribution from usual 3-D turbulence generated by local structural shears on the
background of almost zero vertical gradient of potential temperature (or buoyancy);}
\item{
turbulent entrainment layer at the CBL upper boundary, characterised by essentially stable
stratification with negative (downward) turbulent flux of potential temperature (or buoyancy).}
\end{itemize}

The goal of this paper is to develop the energy and flux budget (EFB) turbulence closure theory for
the surface layer in convective and stably stratified turbulence using budget equations
for turbulent energies and fluxes.
The EFB theory of turbulence closure has been previously developed for stably stratified dry
atmospheric flows \citep{ZKR07,ZKR08,ZKR09,ZKR10,ZKR13,KRZ19}
and for passive scalar transport in stratified turbulence \citep{KRZ21}.
The EFB turbulence closure theory is based on the budget equations for the densities of turbulent
kinetic and potential energies, and turbulent fluxes of momentum and heat.

In agreement with wide experimental evidence \citep{KK78,BB97,OH01,SF01,RK04,SR10,ML10,ML14,L19},
the EFB theory for the stably stratified turbulence \citep{ZKR07,ZKR08,ZKR09,ZKR10,ZKR13,KRZ19}
demonstrates that strong turbulence is maintained
by large-scale shear for any stratification, and the ''critical Richardson number", treated many years as a threshold
between the turbulent and laminar regimes, actually separates two turbulent regimes: the strong turbulence typical
of atmospheric boundary layers and the weak three-dimensional turbulence typical of the free atmosphere,
and characterized by strong decrease in heat transfer in comparison to momentum transfer.

The physical mechanism of self-existence of a stably stratified turbulence is as follows \citep{ZKR13,KRZ21}.
The increase of the vertical gradient of the mean potential temperature (i.e., the increase of the buoyancy)
causes a conversion of turbulent kinetic energy into turbulent potential energy.
On the other hand,  the negative down-gradient vertical turbulent heat flux is decreased
by the counteracting positive non-gradient heat flux that is increased with the
increase of the turbulent potential energy.
The latter is the mechanism of the self-control feedback resulting in
a decrease of the buoyancy.
Due to this feedback, the stably stratified turbulence is maintained up to strongly supercritical stratifications.
The EFB theory have been verified against scarce data from the atmospheric experiments, direct numerical simulations (DNS),
large-eddy simulations (LES) and laboratory experiments relevant to the steady state turbulence regime.

The EFB theory is a sort of the turbulence closure.
Previously, various closure models have been adopted in turbulence and turbulent transport \citep{MY71,MY75,UB05,LE08,CA08,SC08,HL11,D13,RI21}.
Some of the turbulent  closure models for stably stratified atmospheric turbulence
also do not imply a critical Richardson number \citep{MSZ07,GSA07,CCH08,LPR08,SG08,S09,KC09,K10,LK16,L19,CCH20},
see also \citep{OT87}.

In the present study we develop the EFB theory for the surface layers in convective and stably stratified turbulence
which allows us to determine the vertical profiles for all turbulent characteristics.
This paper is organized as follows.
In Section II
we formulate governing equations for the energy and flux budget
turbulence-closure theory for convective and stably stratified turbulence.
In this section we also discuss assumptions used in the EFB theory.
In Section III we develop the EFB theory for surface layers in stratified turbulence
considering the steady-state and homogeneous regime of turbulence.
In Section IV we discuss the EFB theory for
the atmospheric stably stratified boundary-layer turbulence.
In Section V we apply the EFB theory to surface layers in turbulent convection.
Finally, conclusions are drawn in Section VI.

\section{Energy- and flux-budget equations and basic assumptions}
\label{sec:2}

We consider plain-parallel, unstably and stably stratified dry-air flow and employ the budget equations underlying
turbulence-closure theory in the Boussinesq approximation.
We assume that vertical component
of the mean-wind velocity is negligibly small compared to horizontal component, and horizontal gradients of all
properties of the mean flow (the mean velocity and the mean potential temperature) are negligibly small compared to vertical gradients.

In this section we outline the energy and flux budget (EFB) closure theory
based on the budget equations for the density of turbulent
kinetic energy, the intensity of potential temperature fluctuations and turbulent fluxes of momentum and heat.
In our analysis, we use budget equations for the one-point second moments
to develop a mean-field theory. We do not study small-scale structure
of turbulence like intermittency described by high-order moments for turbulent quantities.
We are interested by large-scale long-term dynamics and consider effects in the spatial scales
which are much larger than the integral scale of turbulence
and in timescales which are much longer than the turbulent timescales.

We start with the basic equations of the EFB theory.
The budget equation for the density of turbulent kinetic energy (TKE), $E_{\rm K}=\langle {\bm u}^2 \rangle/2$, reads
\begin{eqnarray}
{DE_{\rm K} \over Dt} + \nabla_z \, \Phi_{\rm K} &=& - \tau_{i z} \, \nabla_z \meanU_i + \beta \, F_z - \varepsilon_{\rm K} ,
 \label{C1}
\end{eqnarray}
where the first term, $- \tau_{i z} \, \nabla_z \meanU_i$, in the right-hand side of Eq.~(\ref{C1}) is the rate
of production of TKE by the vertical gradient of horizontal mean velocity  $\meanUU(z)$,
$D / Dt = \partial /\partial t + \meanUU {\bf \cdot} \bec{\nabla}$  is the convective derivative,
$\tau_{iz} =\langle u_i \, u_z \rangle$ with $i=x, y$ are the off-diagonal components of the Reynolds
stress describing the vertical turbulent flux of momentum, and the angular brackets imply ensemble averaging.
The second term $\beta \, F_z$ in Eq.~(\ref{C1}) describes buoyancy,
$\beta=g/T_\ast$ is the buoyancy parameter, ${\bf g}$  is the gravity acceleration,
$F_z=\langle u_z \, \theta \rangle$ is the vertical component of the turbulent flux of potential temperature,
$\Theta = T (P_\ast / P)^{1-\gamma^{-1}}$ is the potential temperature,
$T$ is the fluid temperature with the reference value $T_\ast$,
$P$ is the fluid pressure with the reference value $P_\ast$
and $\gamma = c_{\rm p}/c_{\rm v}$ is the specific heat ratio.

The potential temperature $\Theta = \overline{\Theta} + \theta$ is characterized by the mean potential
temperature $\overline{\Theta}(z)$ and fluctuations $\theta$,
the fluid velocity $\meanUU + {\bf u}$ is characterized by the mean fluid velocity [which generally includes
the mean-wind velocity $\meanUU^{({\rm w})}(z)=(\meanU_x, \meanU_y, 0)$
and the local three-dimensional mean velocity $\meanUU^{({\rm s})}$ related to the large-scale semi-organised coherent structures in a convective turbulence, and small-scale fluctuations ${\bf u}=(u_x, u_y, u_z)$.

The last term, $\varepsilon_{\rm K} = \nu \, \langle (\nabla_j u_i)^2 \rangle$, in the right-hand side of Eq.~(\ref{C1})
is the dissipation rate of the density of the turbulent kinetic energy, where $\nu$ is the kinematic viscosity
of fluid. The term $\Phi_{\rm K} = \rho_0^{-1} \langle u_z \, p\rangle + (\langle u_z
\, {\bf u}^2 \rangle - \nu \, \nabla_z \langle {\bf u}^2 \rangle)/2$ determines the flux of $E_{\rm K}$,
where the fluid pressure $P = \overline{P} + p$ is characterized by the mean pressure $\overline{P}$
and fluctuations $p$, and $\rho_0$ is the fluid density.

The budget equation for the intensity of potential temperature fluctuations $E_\theta=\langle \theta^2 \rangle/2$ is
\begin{eqnarray}
{D E_\theta \over Dt} + \nabla_z \, \Phi_\theta &=& - F_z \, \nabla_z \overline{\Theta} - \varepsilon_\theta,
 \label{C2}
\end{eqnarray}
where $\Phi_\theta = \left(\langle u_z \, \theta^2 \rangle - \chi \, \nabla_z \langle  \theta^2 \rangle\right)/2$
describes the flux of $E_\theta$ and $\varepsilon_\theta = \chi \, \langle ({\bm \nabla} \theta)^2 \rangle$
is the dissipation rate of the intensity of potential temperature fluctuations $E_\theta$, and $\chi$ is the molecular temperature diffusivity.

The budget equation for the turbulent flux $F_i = \langle u_i \, \theta \rangle$ of potential temperature is given by
\begin{eqnarray}
&&{\partial F_i \over \partial t} + \nabla_z \, {\bm \Phi}_{i}^{({\rm F})}
= - \tau_{iz} \, \nabla_z \overline{\Theta}  + 2 \beta \, E_\theta \, \delta_{i3} - {1 \over \rho_0} \, \langle \theta \, \nabla_i p \rangle
\nonumber\\
&& \qquad- F_z \, \nabla_z \meanU_i - \varepsilon_i^{({\rm F})} ,
\label{C3}
\end{eqnarray}
where $\delta_{ij}$ is the Kronecker unit tensor, ${\bm \Phi}_{i}^{({\rm F})} = \langle u_i \, u_z
\, \theta\rangle - \nu \, \langle \theta \, (\nabla_z u_i) \rangle - \chi\, \langle u_i \, (\nabla_z \theta) \rangle$
determines the flux of $F_i$, and $\varepsilon_i^{({\rm F})} = (\nu + \chi) \, \langle (\nabla_j u_i)
\, (\nabla_j \theta) \rangle$  is the dissipation rate of the turbulent heat flux.
The first term, $- \tau_{iz} \, \nabla_z \overline{\Theta}$,  in the right-hand side of Eq.~(\ref{C3})
contributes to the traditional turbulent flux of potential temperature
which describes the classical gradient mechanism of the turbulent heat transfer.
The second and third terms in the right-hand side
of Eq.~(\ref{C3}) describe a non-gradient contribution to the turbulent flux of
potential temperature.
The budget equation for the vertical turbulent flux $F_z = \langle u_z \, \theta \rangle$ of potential temperature is given by
\begin{eqnarray}
&&{\partial F_z \over \partial t} + \nabla_z \, {\bm \Phi}_{z}^{({\rm F})}
= - 2 E_{z} \, \nabla_z \overline{\Theta} + 2 \beta \, E_\theta - {1 \over \rho_0} \, \langle \theta \, \nabla_z p \rangle
\nonumber\\
&& \qquad- \varepsilon_z^{({\rm F})} ,
\label{CCC3}
\end{eqnarray}
where $E_z=\langle u_z^2 \rangle/2$ is the density of the vertical turbulent kinetic energy.

The budget equation for the off-diagonal components of the Reynolds stress $\tau_{iz} =\langle u_i \, u_z \rangle$ with $i=x, y$ reads
\begin{eqnarray}
{D \tau_{iz} \over Dt} + \nabla_z \, \Phi_{i}^{(\tau)} &=& - 2 \, E_z \, \nabla_z \meanU_i + \beta \, F_i + Q_{iz}
- \varepsilon_{iz}^{(\tau)} ,
 \label{C15}
\end{eqnarray}
where
$\Phi_{i}^{(\tau)}=\langle u_i \, u_z^2 \rangle + \rho_0^{-1} \, \langle p \, u_i\rangle - \nu \, \nabla_z \tau_{iz}$ describes the flux of $\tau_{iz}$,
the tensor $Q_{ij} = \rho_0^{-1} (\langle p \nabla_i u_j\rangle + \langle p \nabla_j u_i\rangle)$
and $\varepsilon_{iz}^{(\tau)}=2 \nu \, \langle (\nabla_j u_i) \, (\nabla_j u_z) \rangle$ is the
molecular-viscosity dissipation rate.

The budget equations for the horizontal and vertical turbulent kinetic energies
$E_\alpha = \langle u_\alpha^2\rangle/2$ can be written as follows:
\begin{eqnarray}
&& {DE_\alpha \over Dt} + \nabla_z \, \Phi_{\alpha} = - \tau_{\alpha z} \, \nabla_z \meanU_\alpha
+ \delta_{\alpha 3} \, \beta \, F_z
\nonumber\\
&& \qquad+ {1 \over 2} Q_{\alpha\alpha}
- \varepsilon_{\alpha},
\label{C4}
\end{eqnarray}
where $\alpha=x,y,z$, the term $\varepsilon_{\alpha} = \nu \, \langle (\nabla_j u_\alpha)^2 \, \rangle$
is the dissipation rate of $E_\alpha$ and
$\Phi_{\alpha}$ determines the flux of $E_\alpha$. Here
$\Phi_{z}= \rho_0^{-1} \langle u_z \, p\rangle + (\langle u_z^3 \rangle - \nu \, \nabla_z \langle u_z^2 \rangle) / 2$
and $\Phi_{x,y}= (\langle u_z \, u_{x,y}^2 \rangle - \nu \, \nabla_z \langle u_{x,y}^2 \rangle) / 2$.
The terms $Q_{\alpha\alpha} = 2 \rho_0^{-1} (\langle p \nabla_\alpha u_\alpha\rangle$ are the diagonal terms of the tensor $Q_{ij}$.
In Eq.~(\ref{C4}) we do not apply the summation convention for the double Greek indices.
Different aspects related to budget equations~(\ref{C1})--(\ref{C4})
have been discussed in a number of publications
\citep{OT87,CM93,KF94,CCH08,ZKR07,ZKR08,ZKR09,ZKR10,ZKR13,KRZ19,KRZ21}.

The energy and flux budget turbulence closure theory
assumes the following.
The characteristic times of variations of the densities of
the turbulent kinetic energies $E_{\rm K}$ and $E_\alpha$,
the intensity of potential temperature fluctuations $E_\theta$,
the turbulent flux $F_i$ of potential temperature and
the turbulent flux  $\tau_{iz}$ of momentum
(i.e., the off-diagonal components of the Reynolds stress)
are much larger than the turbulent timescale.
This allows us to obtain steady-state solutions
of the budget equations~(\ref{C1})--(\ref{C4}).

Dissipation rates of the turbulent kinetic energies $E_{\rm K}$ and $E_\alpha$, the intensity
of potential temperature fluctuations $E_\theta$ and $F_i$ are expressed using
the Kolmogorov  hypothesis, i.e.,
$\varepsilon_{\rm K}=E_{\rm K}/t_{\rm T}$, $\varepsilon_\theta=E_\theta/(C_{\rm p} \, t_{\rm T})$,
and $\varepsilon_i^{({\rm F})}=F_i/(C_{\rm F} \, t_{\rm T})$, where
$t_{\rm T}= \ell_z /E_{z}^{1/2}$ is the turbulent dissipation timescale, $\ell_z$ is the vertical integral scale,
and $C_{\rm p}$  and $C_{\rm F}$ are dimensionless
empirical constants \citep{K41,K42,MY71,MY75,RI21}.
Note also that the dissipation rate of the TKE components $E_\alpha$  (where $\alpha=x, y, z$)
is $\varepsilon_{\alpha} = E_{\rm K}/3t_{\rm T}$. This is because
the main contribution to the rate of dissipation of the TKE components
is from the Kolmogorov viscous scale where turbulence is nearly isotropic,
so that  $\varepsilon_{x} = \varepsilon_{y}= \varepsilon_{z} =E_{\rm K}/3t_{\rm T}$.

The term $\varepsilon_{i}^{(\tau)} = \varepsilon_{iz}^{(\tau)} - \beta \, F_i - Q_{iz}$ in Eq.~(\ref{C15})
is the effective dissipation rate of the off-diagonal components of the Reynolds stress $\tau_{iz}$  \citep{ZKR07,ZKR13,KRZ21},
where $\varepsilon_{iz}^{(\tau)}$ is the molecular-viscosity
dissipation rate of $\tau_{iz}$, that is small because the smallest eddies associated with viscous dissipation
are nearly isotropic \citep{LPR09}.
In the framework of EFB theory, the role of the dissipation rate of $\tau_{iz}$ is assumed to be played by
the combination of terms $- \beta \, F_i - Q_{iz}$, and it is assumed that $\varepsilon_{i}^{(\tau)}
=\tau_{iz}/(C_\tau \, t_{\rm T})$, where $C_\tau$ is the effective-dissipation time-scale empirical constant
for stably stratified turbulence \citep{ZKR07,ZKR13,KRZ21}, while for a convective turbulence $C_\tau$
is a function of the flux Richardson number (see Sect.~\ref{sec:5}).

The effective dissipation rate assumption has been justified by
Large Eddy Simulations (see Fig.~1 in \cite{ZKR13}),
where LES data by \cite{ES04,EZ06}  have been used for the two types of atmospheric boundary layer: “nocturnal stable”
(with essentially negative buoyancy flux at the surface and neutral stratification in
the free flow) and “conventionally neutral” (with a negligible buoyancy flux at the
surface and essentially stably stratified turbulence in the free flow).
The effective dissipation rate assumption was based on
our prior analysis of the Reynolds stress equation in the ${\bm k}$-space using the $\tau$ -approximation
(see \cite{EKRZ02,EKRZ06}).
Remarkably, the effective dissipation assumption directly yields
the familiar down-gradient formulation of the vertical turbulent flux of momentum
[see Eq.~(\ref{A3}) below], that is well-known result which is valid for any turbulence
with a non-uniform mean velocity field.

Note that the diagonal and off-diagonal components of the Reynolds stress have different physical meaning.
The diagonal components of the Reynolds stress describe turbulent kinetic energy components.
They have  the Kolmogorov spectrum $\propto k^{-5/3}$,
that is related to the direct energy cascade.
The latter is the main reason for turbulent viscosity and turbulent diffusivity.
The off-diagonal components of the Reynolds stress are related to the tangling mechanism
of generation of anisotropic velocity fluctuations. They have different spectrum $\propto k^{-7/3}$
\citep{L67,WC72,SV94,IY02}.
The off-diagonal components of the Reynolds stress are determined by spatial derivatives
of the mean velocity field.
The diagonal components of the Reynolds stress are much larger than the off-diagonal components.

We assume that the term $\rho_0^{-1} \, \langle \theta \,
\nabla_z p \rangle$ in Eq.~(\ref{CCC3}) for the vertical turbulent flux of potential
temperature is parameterised so that
$\beta \, \langle \theta^2
\rangle  - \rho_0^{-1} \, \langle \theta \,
\nabla_z p \rangle = 2 C_\theta \, \beta \, E_\theta $,
with the positive dimensionless empirical constant $C_\theta$
which is less than 1.
This assumption has been justified by
Large Eddy Simulations (see Fig.~2 in \cite{ZKR13}),
where LES data by \cite{ES04,EZ06}  have been used for the two types of atmospheric boundary layer: “nocturnal stable” and “conventionally neutral”.
In addition, this assumption  has been justified analytically
(see Appendix A in \cite{ZKR07}).

\section{The EFB theory for surface layers in stratified turbulence}
\label{sec:3}

In this section we develop the EFB theory for surface layers in convective and stably stratified turbulence.
We use the down-gradient formulation of the vertical turbulent flux of momentum
which follows from Eq.~(\ref{C15}), i.e., the turbulent fluxes of the momentum are
\begin{eqnarray}
&& \tau_{iz}= - K_{\rm M} \, \nabla_z \meanU_i ,  \quad i=x, y,
\label{A3}\\
&& K_{\rm M} = 2 C_\tau \, t_{\rm T} \, E_z  = 2 C_\tau \, \ell_z \, E_z^{1/2},
\label{A4}
\end{eqnarray}
where $K_{\rm M}$ is the turbulent viscosity, $t_{\rm T}= \ell_z /E_{z}^{1/2}$ is the turbulent dissipation timescale, $\ell_z$ is the vertical integral scale
and $E_{z}$ is the vertical turbulent kinetic energy.
The production rate, $\Pi_{\rm K} = - \tau_{i z} \, \nabla_z \meanU_i$ of the turbulent kinetic energy by the vertical gradient of horizontal mean velocity [see Eq.~(\ref{C1})]
can be rewritten by means of Eq.~(\ref{A3}) as $\Pi_{\rm K} = - (\tau_{x z} \, \nabla_z \meanU_x + \tau_{y z} \, \nabla_z \meanU_y)= K_{\rm M} \, S^2$,
where $S^2 = (\nabla_z \meanU_x)^2 + (\nabla_z \meanU_y)^2$ is the squared mean velocity shear caused by the horizontal mean wind velocity.

The steady-state version of the budget equations for the density of turbulent kinetic energy $E_{\rm K}=\langle {\bm u}^2 \rangle/2$ reads
\begin{eqnarray}
\nabla_z \, \Phi_{\rm K} = K_{\rm M} S^2 + \beta \, F_z - {E_{\rm K} \over t_{\rm T}} ,
 \label{A1}
\end{eqnarray}
where the dissipation rate $\varepsilon_{\rm K}$ of the turbulent kinetic energy is expressed using
the Kolmogorov  hypothesis, $\varepsilon_{\rm K}=E_{\rm K}/t_{\rm T}$.
We stress that all results obtained in the present study are mainly valid for temperature  stratified turbulence (convective turbulence or
stably stratified turbulence), where fluctuations of the vertical velocity $u_z$ depend on the buoyancy, $\beta \, F_z$.
Since for temperature  stratified turbulence, $\rho_0^{-1} \langle u_z \, p\rangle$ and $\langle u_z \, {\bf u}^2 \rangle$ do depend on
the buoyancy, the third-order moments $\Phi_{\rm K}$ should depend on buoyancy.
We assume that the vertical gradient $\nabla_z \, \Phi_{\rm K}$ of the flux of $E_{\rm K}$ is determined by the buoyancy, i.e.,
$\nabla_z \Phi_{\rm K} = - C_\Phi \, \beta \, F_z$, where $C_\Phi$ is the dimensionless empirical constant.
The justification of this assumption for a convective turbulence has been performed in  Ref.~\cite{ZRK21},
where experimental data
obtained from meteorological observations at the Eureka
station (located in the Canadian territory of Nunavut) in conditions
of the long-lived convective boundary layer typical of
the Arctic summer have been used for validation of the assumption $\nabla_z \Phi_{\rm K} = - C_\Phi \, \beta \, F_z$
(see the right panel in Fig.~1 in  Ref.~\cite{ZRK21}).
Turbulent fluxes were calculated directly from
the measured velocity and temperature fluctuations.
In these  meteorological observations warming of the
convective layer from the surface is balanced by pumping of colder
air into the layer via the general-circulation mechanisms.
Note also that no principal contradictions
have been found between the available data from observations at
mid- or low latitudes and the data from Eureka
\citep{GP18}.

Using the expression,
\begin{eqnarray}
\tau = \left(\tau_{xz}^2 + \tau_{yz}^2\right)^{1/2} = K_{\rm M} \, S  \equiv u_\ast^2,
\label{PTT10}
\end{eqnarray}
Eq.~(\ref{A1}) is reduced by simple algebraic calculations to a nonlinear equation for the vertical profile of the normalized TKE, $\tilde E_{\rm K}(\tilde Z) = E_{\rm K}(\tilde Z)/E_{\rm K0}$ as
\begin{eqnarray}
\tilde E_{\rm K}^2 + \tilde Z \, \tilde E_{\rm K}^{1/2} - 1 =0 ,
 \label{A6}
\end{eqnarray}
where the normalised height $\tilde Z = \ell_z / (C_\ast \, L)$, $E_{\rm K0} = u_\ast^2 / (2 C_\tau \, A_z)^{1/2}$,
 $C_\ast^{-1} = (1 + C_\Phi) \, (2 C_\tau)^{3/4} A_z^{1/4}$, $\,A_z=E_{z}/E_{\rm K}$ is the vertical share of TKE, $u_\ast$ is the local ($z$-dependent) friction velocity,
 and $L$ is the local Obukhov length defined as
\begin{eqnarray}
L= - {\tau^{3/2} \over \beta \, F_z} ,
\label{T10}
\end{eqnarray}
and $F_z$ is the local vertical turbulent flux  of potential temperature.
For stably stratified turbulence, the vertical turbulent flux $F_z$ of potential temperature is negative, and the local Obukhov length $L$ is positive. For stably stratified turbulence ($\tilde Z>0$),  Eq.~(\ref{A6})  has two asymptotic solutions:

(i) for a lower part ($ \tilde Z  \ll 1$) of the surface  layer, Eq.~(\ref{A6}) yields
\begin{eqnarray}
\tilde E_{\rm K} = 1 - {\tilde Z \over 2} +{\tilde Z^2 \over 8}  ,
 \label{TA7a}
\end{eqnarray}

(ii) for an upper part ($ \tilde Z  \gg 1$) of the surface  layer, it is
\begin{eqnarray}
\tilde E_{\rm K} = \tilde Z^{-2} \, \left(1 - 2  \tilde Z^{-4} \right) .
 \label{TA7b}
\end{eqnarray}

In the framework of the EFB theory of surface layers, we use
the same definition~(\ref{T10}) for $L$ in convective turbulence as well,
where the vertical  turbulent flux $F_z$ of potential temperature is positive, and $L$ is negative.
Equation~(\ref{A6})  for the surface layer in convective turbulence ($\tilde Z<0$)  reads
\begin{eqnarray}
\tilde E_{\rm K}^2 - |\tilde Z| \, \tilde E_{\rm K}^{1/2} - 1 =0 ,
 \label{MA6}
\end{eqnarray}
and it has two asymptotic solutions:

(i) for a lower part ($ |\tilde Z|  \ll 1$) of the surface convective layer, Eq.~(\ref{MA6}) yields
\begin{eqnarray}
E_{\rm K} = E_{\rm K0} \, \left(1 + {1 \over 2} \,  |\tilde Z|\right) ,
 \label{A7}
\end{eqnarray}

(ii) for an upper part ($ |\tilde Z|  \gg 1$) of the surface convective layer, the balance of the first and the second terms
in Eq.~(\ref{A6}) yields $\tilde E_{\rm K} =  \tilde Z^{2/3}$, i.e.,
\begin{eqnarray}
E_{\rm K} = E_{\rm K0} \, \tilde Z^{2/3} .
 \label{A8}
\end{eqnarray}
Note that as follows from the definition of $\tilde Z=\ell_z / (C_\ast \, L)$, the ratio $z/L$
for convective turbulence is
\begin{eqnarray}
{z \over L} = {\tilde Z \over \kappa_0 \,  (1 + C_\Phi)} ,
\label{LL10}
\end{eqnarray}
where $\ell_z=C_\ell \, z$ with  $C_\ell = \kappa_0 \, (2 C_\tau)^{-3/4} A_z^{-1/4}$ and $\kappa_0=0.4$ is the von Karman constant.
Note that generally, Eq.~(\ref{LL10})  can be valid also for arbitrary $z$,
but in this case $C_\ell $ should be a function of height (see Section~\ref{sec:5}).

\begin{figure}
\centering
\includegraphics[width=8.0cm]{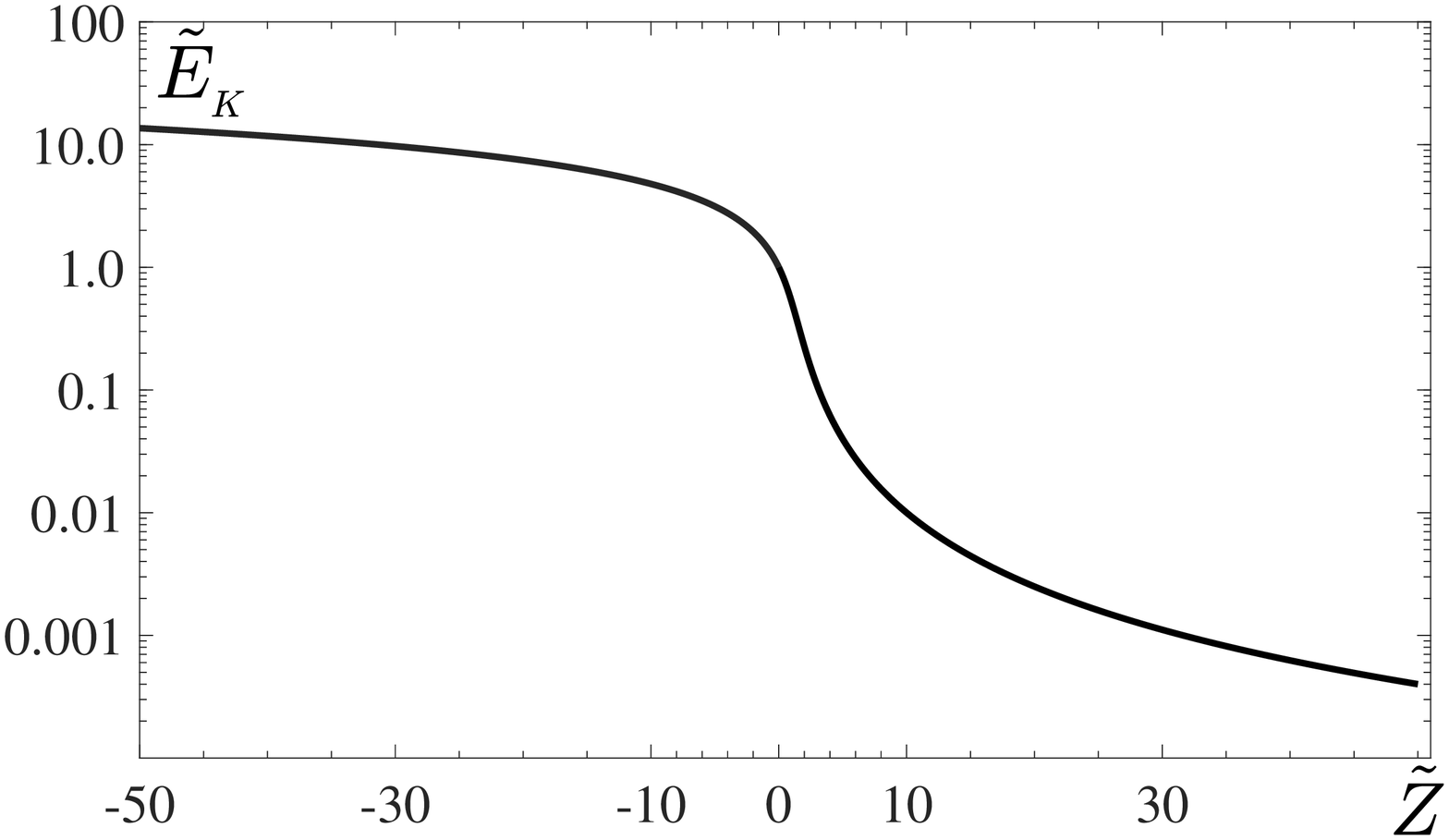}
\caption{\label{Fig1} The normalized turbulent kinetic energy $\tilde E_{\rm K} = E_{\rm K}/E_{\rm K0}$
versus $\tilde Z$ for convective and stably stratified turbulence.
}
\end{figure}

 In Fig.~\ref{Fig1} we show a numerical solution of Eq.~(\ref{A6}). In particular, in Fig.~\ref{Fig1} we plot the normalized
 turbulent kinetic energy $\tilde E_{\rm K} = E_{\rm K}/E_{\rm K0}$
versus $\tilde Z$  for convective ($\tilde Z<0$) and stably stratified ($\tilde Z>0$) turbulence.
This numerical solution is in a good agreement with the above asymptotic solutions for convective and stably stratified turbulence.

Now we define the flux Richardson number as
\begin{eqnarray}
{\rm Ri}_{\rm f} = - {\beta \, F_z \over K_{\rm M} S^2} ,
\label{MML13}
\end{eqnarray}
so that for stably stratified turbulence, ${\rm Ri}_{\rm f}$ is positive and varies from 0 to the limiting value
$R_\infty=0.2$ at very large gradient Richardson number ${\rm Ri} \gg 1$. Here the gradient Richardson number,
${\rm Ri}$, is defined as
\begin{eqnarray}
{\rm Ri} = {N^2 \over S^2} ,
\label{C12}
\end{eqnarray}
where $N^2 = \beta \, \nabla_z \overline{\Theta}$  and $N$ is the Brunt-V\"{a}is\"{a}l\"{a} frequency.
In the framework of  the EFB theory of surface layers, we use the same definition~(\ref{MML13}) for
the flux Richardson number in turbulent convection, so that ${\rm Ri}_{\rm f}$ is negative in turbulent convection,
and its absolute value is not limited and can be large.

Equations~(\ref{T10}) and~(\ref{MML13})  allow us  to relate the turbulent viscosity $K_{\rm M}$
with the flux Richardson number ${\rm Ri}_{\rm f}$ as \cite{KRZ21}
\begin{eqnarray}
K_{\rm M} = {\rm Ri}_{\rm f} \, \tau^{1/2} \, L  ,
\label{TT10}
\end{eqnarray}
where $\tau$ is given by Eq.~(\ref{PTT10}).
Using Eqs.~(\ref{A4}) and~(\ref{TT10}), we rewrite
the flux Richardson number as
\begin{eqnarray}
{\rm Ri}_{\rm f} = (1 + C_\Phi)^{-1} \, \tilde Z \, \tilde E_{\rm K}^{1/2} .
\label{ML13}
\end{eqnarray}
Equations~(\ref{PTT10})  and~(\ref{TT10})
allow us  to relate the large-scale shear $S$ with the flux Richardson number as
\begin{eqnarray}
S = {\tau^{1/2} \over L \, {\rm Ri}_{\rm f}}  .
\label{AA11}
\end{eqnarray}
Using Eqs.~(\ref{A4}),  we  rewrite Eq.~(\ref{A1}) as the dimensionless ratio
\begin{eqnarray}
\left({E_{\rm K} \over \tau}\right)^2 = { 1 - {\rm Ri}_{\rm f} \,  (1 + C_\Phi) \over 2 C_\tau \, A_z} .
\label{C9}
\end{eqnarray}
In addition, by means of Eqs.~(\ref{A4}), (\ref{TT10}) and~(\ref{C9}),  we obtain the normalized vertical integral scale $\ell_z$
as the function of the flux Richardson number:
\begin{eqnarray}
{\ell_z  \over L}  = {(2 C_\tau)^{-3/4}  \, A_z^{-1/4} \, {\rm Ri}_{\rm f} \over  \left[1 - {\rm Ri}_{\rm f}\,  (1 + C_\Phi) \right]^{1/4}} ,
\label{FF8}
\end{eqnarray}
where $1 - {\rm Ri}_{\rm f}\,  (1 + C_\Phi) > 0$. This condition implies that $C_\Phi < R_\infty^{-1}-1$.
For stably stratified turbulence $R_\infty=0.2$, so that $C_\Phi < 4$.
Thus, the normalised height $\tilde Z = \ell_z / (C_\ast \, L)$ as the function of the flux Richardson number reads
\begin{eqnarray}
\tilde Z = {{\rm Ri}_{\rm f}\,  (1 + C_\Phi)  \over  \left[1 - {\rm Ri}_{\rm f}\,  (1 + C_\Phi) \right]^{1/4}} .
\label{FFF8}
\end{eqnarray}
Note also that using Eq.~(\ref{FFF8}) we can rewrite Eq.~(\ref{FF8})  as
\begin{eqnarray}
{\ell_z  \over L}  = {C_\ell \, \tilde Z \over \kappa_0 \, (1 + C_\Phi)} .
\label{MFF8}
\end{eqnarray}

Since convective turbulence is essentially different from stably stratified turbulence, the behaviour of the flux Richardson number
${\rm Ri}_{\rm f} \propto \, \tilde Z \, \tilde E_{\rm K}^{1/2}$ is also different for these two kinds of turbulence.
In particular, in convection both, the buoyancy and large-scale shear produce convective turbulence, so that
the flux Richardson number can be enough large.
Contrary, in stably stratified turbulence the large-scale shear produces turbulence,
while the buoyancy decreases TKE, so that the flux Richardson number is limited by
some value, ${\rm R}_\infty=0.2$.
However, in the presence of internal gravity waves the maximum
value of the flux Richardson number can be larger in several times
in comparison with the case without waves \citep{ZKR09,KRZ19}.

Equations~(\ref{A6}) and~(\ref{ML13})  yield the normalized
turbulent kinetic energy $\tilde E_{\rm K} = E_{\rm K}/E_{\rm K0}$ as a function of
the flux Richardson number as
\begin{eqnarray}
\tilde E_{\rm K} =\left[ 1-  (1 + C_\Phi) \, {\rm Ri}_{\rm f}\right]^{1/2}  .
\label{FML13}
\end{eqnarray}
This implies that the normalized turbulent kinetic energy $\tilde E_{\rm K}$ in stably stratified turbulence decreases
up to the minimum value
\begin{eqnarray}
\tilde E_{\rm K}^{\rm min}=[1-  (1 + C_\Phi) \, {\rm R}_\infty]^{1/2} .
\label{FML15}
\end{eqnarray}
As follows from Eq.~(\ref{ML13}), the function $\tilde Z \, \tilde E_{\rm K}^{1/2} \leq (1 + C_\Phi) \, {\rm R}_\infty$,
so that the maximum value of the height $\tilde Z^{\rm max}$ in stably stratified turbulence is
\begin{eqnarray}
\tilde Z^{\rm max}  = {{\rm R}_\infty\,  (1 + C_\Phi)  \over  \left[1 - {\rm R}_\infty\,  (1 + C_\Phi) \right]^{1/4}} .
\label{FML14}
\end{eqnarray}
Since in convective turbulence, the flux Richardson number is negative,
the normalized turbulent kinetic energy, $\tilde E_{\rm K} = [1+ (1 + C_\Phi) \, |{\rm Ri}_{\rm f}|]^{1/2}$,
increases with the flux Richardson number.

Now we derive expression for the turbulent Prandtl number,
${\rm Pr}_{_{\rm T}} = K_{\rm M} / K_{\rm H}$.
To this end, we use the steady-state versions of Eqs.~(\ref{C2}) and~(\ref{CCC3}):
\begin{eqnarray}
&& F_{z} \, \nabla_z \overline{\Theta} + {E_\theta \over C_{\rm p} \, t_{\rm T}} = 0 ,
 \label{MC2}\\
&& 2 E_{z} \, \nabla_z \overline{\Theta} - 2 C_\theta\, \beta \, E_\theta + {F_z \over C_{\rm F} \, t_{\rm T}} = 0 .
 \label{MC3}
\end{eqnarray}
Equations~(\ref{MC2})--(\ref{MC3}) and the expression for the turbulent heat flux,
$F_{z} = - K_{\rm H} \, \nabla_z \overline{\Theta}$, yield the turbulent heat conductivity
$K_{\rm H}$ as
\begin{eqnarray}
K_{\rm H} = 2 C_{\rm F} \, t_{\rm T} \, E_z \left[1 + {C_\theta \, C_{\rm p} \,t_{\rm T} \,\beta F_z
\over E_z}\right] .
\label{MCC1}
\end{eqnarray}
By means of Eq.~(\ref{A1}) for TKE,
\begin{eqnarray}
E_{\rm K}=K_{\rm M} \, S^2 \, t_{\rm T} \,  [1 - {\rm Ri}_{\rm f}\,  (1 + C_\Phi)] ,
\label{DCC1}
\end{eqnarray}
and Eq.~(\ref{MML13}) for ${\rm Ri}_{\rm f}$, we derive the identity for the dimensionless ratio as
\begin{eqnarray}
{\beta \, F_z \, t_{\rm T} \over E_z} = - {{\rm Ri}_{\rm f} \over A_z \left[1 - {\rm Ri}_{\rm f} \,  (1 + C_\Phi) \right]} .
\label{DD5}
\end{eqnarray}
Thus, Eqs.~(\ref{A4}), (\ref{MCC1}) and~(\ref{DD5}) yield the turbulent Prandtl number,
${\rm Pr}_{_{\rm T}} = K_{\rm M} / K_{\rm H}$ as
\begin{eqnarray}
{\rm Pr}_{_{\rm T}} = {\rm Pr}_{_{\rm T}}^{(0)} \,  \left[1 - { C_\theta \, C_{\rm p} \,{\rm Ri}_{\rm f}
\over A_z \left[1 - {\rm Ri}_{\rm f} \,  (1 + C_\Phi)\right]}
\right]^{-1} ,
\label{MDD5}
\end{eqnarray}
where ${\rm Pr}_{_{\rm T}}^{(0)} = C_\tau / C_{\rm F}$ is the turbulent Prandtl number for a non-stratified turbulence
when the mean potential temperature gradient vanishes.
The gradient Richardson number Ri and the flux Richardson number ${\rm Ri}_{\rm f}$
are related as ${\rm Ri} = {\rm Ri}_{\rm f} \, {\rm Pr}_{_{\rm T}}$.

Using Eqs.~(\ref{MC2})--(\ref{MC3}), we determine the level of temperature fluctuations characterised
by the dimensional ratio $E_\theta/ \theta_\ast^2$ as
\begin{eqnarray}
{E_\theta \over \theta_\ast^2} = C_{\rm p} \, (2 C_\tau \, A_z)^{-1/2} \,   \, {\rm Pr}_{_{\rm T}} \,  \tilde E_{\rm K}^{-1}  ,
 \label{AA9}
\end{eqnarray}
where $\theta_\ast = |F_z|/u_\ast = u_\ast^2 / \beta \, |L|$.
Equation~(\ref{AA11}), and expressions for the friction velocity, $u_\ast^2=K_{\rm M} \, S $, and the turbulent heat flux,
$F_{z} = - K_{\rm H} \, \nabla_z \overline{\Theta}$, yield the vertical gradient of the mean potential temperature as
\begin{eqnarray}
\nabla_z \overline{\Theta} ={ \theta_\ast \,  {\rm Pr}_{_{\rm T}}  \over |L|  \, {\rm Ri}_{\rm f} } .
 \label{FFA12}
\end{eqnarray}

The steady-state version of Eq.~(\ref{C3}) for homogeneous turbulence yields the horizontal turbulent flux $F_i$ of potential temperature:
\begin{eqnarray}
F_i = - C_{\rm F} \, t_{\rm T} \, \left(1 + {\rm Pr}_{_{\rm T}}\right) \, F_z \, \nabla_z \meanU_i , \quad i=x, y .
\label{CC8}
\end{eqnarray}
Since in convective turbulence the vertical turbulent flux $F_z$ is positive,
the horizontal turbulent flux $F_i$ of potential temperature
is directed opposite to the wind velocity $\meanU_i$, i.e., Eq.~(\ref{CC8}) describes the counter-wind
horizontal turbulent flux in convective turbulence.
Contrary, in a stably stratified turbulence the vertical turbulent flux $F_z$ is negative,
so that Eq.~(\ref{CC8}) determines the co-wind horizontal turbulent flux.

The physics of the counter-wind turbulent flux in a convective turbulence is as follows  \citep{KRZ21}.
In horizontally homogeneous, convective turbulence with a large-scale shear velocity (e.g., directed along the $x$-axis),
the mean shear velocity $\meanU_x$ increases with increasing height,
while the mean potential temperature $\overline{\Theta}$ decreases with height.
Uprising fluid particles produce positive fluctuations of potential temperature ($\theta>0$)
since $\partial \theta/\partial t \propto - ({\bm u} \cdot
\bec{\nabla}) \overline{\Theta}$, and negative fluctuations of horizontal velocity ($u_x<0$)
since $\partial u_x/\partial t \propto - ({\bm u} \cdot
\bec{\nabla}) \meanU_x$. It results in negative horizontal temperature flux, $u_x \, \theta<0$.
Similarly, sinking fluid particles cause negative fluctuations of potential temperature ($\theta<0$),
and positive fluctuations of horizontal velocity ($u_x>0$), that implies
negative horizontal temperature flux
$u_x \, \theta<0$. Therefore, the net horizontal turbulent flux is negative
($\langle u_x \, \theta\rangle<0$) even for a zero horizontal mean temperature gradient.
This is the counter-wind turbulent flux of potential temperature that
results in modification of the potential-temperature flux by
the non-uniform velocity field.

Let us find dependence of the horizontal turbulent flux $F_i$ of potential temperature
on the flux Richardson number. To this end we use the identity,
\begin{eqnarray}
\left(S \, t_{\rm T}\right)^2 ={1 \over 2 \, C_\tau  \,  A_z \, [1 - {\rm Ri}_{\rm f}\,  (1 + C_\Phi)]},
\label{MD5}
\end{eqnarray}
that is derived by means of Eqs.~(\ref{A4}), (\ref{PTT10}) and~(\ref{C9}).
Therefore, the ratio of the horizontal and vertical turbulent fluxes of potential temperature, $F_x/F_z$, is given by
\begin{eqnarray}
&& {F_x \over F_z}= - C_{\rm F}  \, \left(1 + {\rm Pr}_{_{\rm T}}\right) \, \left(2 \, C_\tau  \,  A_z\right)^{-1/2}
\nonumber\\
&& \qquad\times   \Big[1 - {\rm Ri}_{\rm f}\,  (1 + C_\Phi)\Big]^{-1/2} .
\label{MCC8}
\end{eqnarray}

Most of the results obtained in this section
depend on the vertical share of TKE, $A_z \equiv E_z / E_{\rm K}$, which is determined below.
The mean shear velocity $\meanU_x(z)$ produces the energy of longitudinal velocity fluctuations $E_x$,
which in turns feeds the transverse $E_y$  and the vertical $E_z$ components
of turbulent kinetic energy. The inter-component energy exchange term $Q_{\alpha\alpha}$ in
the right-hand side of Eq.~(\ref{C4})
is traditionally parameterized through the ''return-to-isotropy" hypothesis \citep{R51}.
On the other hand, the temperature stratified turbulence is usually anisotropic,
and the inter-component energy exchange term $Q_{\alpha\alpha}$
should depend on the flux Richardson number ${\rm Ri}_{\rm f}$.

Here we adopt the following model for the inter-component energy exchange term $Q_{\alpha\alpha}$
which generalizes the ''return-to-isotropy" hypothesis to the case of the convective and stably stratified turbulence.
We use the normalised flux Richardson number ${\rm Ri}_{\rm f}/R_\infty$ that is varying from 0
for a non-stratified turbulence to 1 for a strongly stratified turbulence,
where the limiting value of the flux Richardson number, ${\rm R}_\infty \equiv
{\rm Ri}_{\rm f}|_{_{{\rm Ri} \to \infty}}$, is defined for very strong stratifications
when the gradient Richardson number ${\rm Ri} \to \infty$.
The model for the inter-component energy exchange term $Q_{\alpha\alpha}$ is described by
\begin{eqnarray}
Q_{xx} = - {2(1 + C_{\rm r}) \over t_{\rm T}} \,\left( E_x - {1 \over 3} \, E_{\rm int}\right) ,
\label{MAP1}
\end{eqnarray}
\begin{eqnarray}
Q_{yy} = - {2(1 + C_{\rm r}) \over t_{\rm T}} \,\left(E_y - {1 \over 3} \, E_{\rm int}\right) ,
\label{MAP2}
\end{eqnarray}
\begin{eqnarray}
Q_{zz} = - {2(1 + C_{\rm r}) \over t_{\rm T}} \,\left(E_z - E_{\rm K}
+ {2 \over 3} \, E_{\rm int}\right) ,
\label{MAP3}
\end{eqnarray}
where
\begin{eqnarray}
E_{\rm int} = E_{\rm K} \, \left[1 - {{\rm Ri}_{\rm f} \over R_\infty} \, \left({C_{\rm r} \over 1 + C_{\rm r}}\right)\right],
\label{MAP4}
\end{eqnarray}
and $C_{\rm r}$ is the dimensionless
empirical constant. When ${\rm Ri}_{\rm f}=0$,
Eqs.~(\ref{MAP1})--(\ref{MAP4}) describe the ''return-to-isotropy" hypothesis \citep{R51}.
To derive equation for the vertical share of TKE
in a stratified turbulence, we use the steady-state version of
Eq.~(\ref{C4}) for vertical TKE $E_z$ as
\begin{eqnarray}
\nabla_z \, \Phi_z  = \beta \, F_z  + {1 \over 2} Q_{zz} - {E_{\rm K} \over 3 t_{\rm T}} .
\label{APC4}
\end{eqnarray}
We assume that the vertical gradient $\nabla_z \, \Phi_z$ of the flux of $E_z$ is determined by the buoyancy, i.e.,
$\nabla_z \Phi_z = - C_z \, \beta \, F_z$, where $C_z$ is the dimensionless empirical constant.
The justification of this assumption for a convective turbulence has been performed in  Ref.~\cite{ZRK21},
where experimental data obtained from meteorological observations at the Eureka
station have been used for validation of this assumption
(see the left panel in Fig. 1 in Ref.~\cite{ZRK21}).
Thus, by means of  Eqs.~(\ref{DD5}) and~(\ref{MAP3})--(\ref{APC4}), we determine
the vertical share of TKE $A_z \equiv E_z / E_{\rm K}$ as a function
of the flux Richardson number:
\begin{eqnarray}
A_z({\rm Ri}_{\rm f})  &=&  A_z^{(0)}   - {\rm Ri}_{\rm f} \,  \left[ {(1-3 A_z^{(0)}) \, (1 + C_z)\over 1 - {\rm Ri}_{\rm f}\,  (1 + C_\Phi)}-  {2 A_z^{(0)} \over {\rm R}_\infty} \right] .
\nonumber\\
\label{MRC11}
\end{eqnarray}
According to Eq.~(\ref{MRC11}), the vertical share  $A_z$ of TKE for a non-stratified turbulence
is $(A_z)_{_{{\rm Ri} \to 0}} \equiv A_z^{(0)} = C_{\rm r}/3(1+C_{\rm r})$.
Usually in surface layers in convective turbulence, $|{\rm Ri}_{\rm f}| \ll |{\rm R}_\infty|$.
This implies that the vertical share of TKE in a convective turbulence
is given by
\begin{eqnarray}
A_z({\rm Ri}_{\rm f})  =  A_z^{(0)}   + (1-3 A_z^{(0)}) \, {|{\rm Ri}_{\rm f}| \, (1 + C_z)\over 1 + |{\rm Ri}_{\rm f}| \,  (1 + C_\Phi)} .
\label{MRC12}
\end{eqnarray}
In convective turbulence for large $|{\rm Ri}_{\rm f}| \gg 1$, the vertical share of TKE $A_z \to 1$ \citep{ZRK21}.
This condition yields
\begin{eqnarray}
{1 + C_z \over 1 + C_\Phi} = {1 -  A_z^{(0)}  \over  1-3 A_z^{(0)}} .
\label{MMRC12}
\end{eqnarray}
Substituting Eq.~(\ref{MMRC12}) into Eq.~(\ref{MRC12}), we obtain that the vertical share of TKE in a stably stratified turbulence is
\begin{eqnarray}
A_z({\rm Ri}_{\rm f})  &=&  A_z^{(0)}   - {\rm Ri}_{\rm f} \,  \left[{1- A_z^{(0)} \over (1 + C_\Phi)^{-1} - {\rm Ri}_{\rm f}}
-  {2 A_z^{(0)} \over {\rm R}_\infty} \right]  ,
\nonumber\\
\label{MMRC11}
\end{eqnarray}
while in a convective turbulence (where $|{\rm Ri}_{\rm f}| \ll |{\rm R}_\infty|$ and ${\rm Ri}_{\rm f}<0$), the vertical share of TKE is
\begin{eqnarray}
A_z({\rm Ri}_{\rm f})  =  A_z^{(0)}   + {\Big(1- A_z^{(0)} \Big)  \,  |{\rm Ri}_{\rm f}| \over (1 + C_\Phi)^{-1} + |{\rm Ri}_{\rm f}|} .
\label{MCRC11}
\end{eqnarray}
Note that Eqs.~(\ref{MAP1})--(\ref{MAP4}) describes a simple generalization of
the ''return-to-isotropy" hypothesis \citep{R51}.
These equations affect only Eq.~(\ref{MMRC11})
for the dependence of the vertical share of TKE
on the flux Richardson number, $A_z({\rm Ri}_{\rm f})$.
This function is the most unknown in observations.

When turbulence is isotropic in the horizontal plane, the horizontal shares of TKE are $A_x=A_y=1 - A_z$. This yields the horizontal components of TKE as
\begin{eqnarray}
E_x=E_y  = {1 \over 2} \, E_{\rm K} \, (1 - A_z) ,
\label{R11}
\end{eqnarray}
where $A_x=E_x/E_{\rm K}$ and $A_y=E_y/E_{\rm K}$.
When turbulence is anisotropic in the horizontal plane,  the inter-component energy exchange term $Q_{\alpha\alpha}$
and the horizontal shares of TKE are given in Appendix.

Let us consider stably stratified turbulence. Neglecting the term $\nabla_z \, \Phi_{\rm K}$ in Eq.~(\ref{A1}),
we rewrite this equation as ${\rm Ri}_{\rm f}^{-1} = 1 - \varepsilon_{\rm K}/ \beta \, F_z$,
where we use the definition~(\ref{MML13}) for the flux Richardson number.
By means of  this equation and  the expressions for the squared Brunt-V\"{a}is\"{a}l\"{a} frequency, $N^2 = \beta \, \nabla_z \overline{\Theta}$,  and the turbulent heat flux, $F_{z} = - K_{\rm H} \, \nabla_z \overline{\Theta}$,
we obtain equation for the turbulent heat conductivity $K_{\rm H}$ as
\begin{eqnarray}
K_{\rm H} = \left({\rm Ri}_{\rm f}^{-1} - 1\right)^{-1} \, {\varepsilon_{\rm K} \over N^2} .
\label{TL1}
\end{eqnarray}
In very strong stable stratification, the gradient Richardson number
admits a limit Ri $\to \infty$ and the flux Richardson number ${\rm Ri}_{\rm f} \to 0.2$.
This implies that the turbulent heat conductivity for a very strong stable stratification
$K_{\rm H} \approx 0.25 \, \varepsilon_{\rm K} / N^2$.
This is a well-known Cox-Osborn equation \citep{OC72,OS80}
that plays an important role in Physical Oceanography.

\section{The atmospheric stably stratified boundary-layer turbulence}

In view of the applications of the obtained results to the atmospheric stably stratified boundary-layer turbulence,
we outline below the useful in modelling theoretical relationships \citep{ZKR13,KRZ21}.
It is known that the wind shear in stably stratified turbulence has two asymptotic results:
(i) $S=\tau^{1/2} / (\kappa_0 \, z)$ at $\varsigma \ll 1$, which describes the log-profile for the mean velocity, and
(ii) $S=\tau^{1/2} / ({\rm R}_\infty \, L)$ when $\varsigma \gg 1$. The latter result follows from Eq.~(\ref{AA11}),
where $\varsigma=\int_0^z \, dz'/L(z')$ is the dimensionless height based on the local Obukhov length scale
$L(z)$, and $\kappa_0=0.4$ is the von Karman constant.
For surface layer in stably stratified turbulence  (defined as the lower layer
which is 10 \% of the turbulent boundary layer), the Obukhov length scale $L$ is independent of $z$ and
the dimensionless height $\varsigma=z/L$.
Interpolating these two asymptotic results, we obtain that the wind shear $S(\varsigma)$ can be written as
\begin{eqnarray}
S = {\tau^{1/2} \over L} \, \left(R_\infty^{-1} +  {1 \over \kappa_0 \, \varsigma} \right) . 
\label{APPF1}
\end{eqnarray}
The latter allows us to get the vertical profile of the turbulent viscosity $K_{\rm M}(\varsigma) = \tau/S$ as
\begin{eqnarray}
K_{\rm M} = \tau^{1/2} \, L \, \, {\kappa_0 \, \varsigma  \over 1 + R_\infty^{-1}
\, \kappa_0 \, \varsigma} .
\label{F1}
\end{eqnarray}
Using  Eqs.~(\ref{TT10}) and~(\ref{F1}), we arrive at the expression for the vertical profile 
of the flux Richardson number ${\rm Ri}_{\rm f}(\varsigma)$ as
\begin{eqnarray}
{\rm Ri}_{\rm f}= {\kappa_0 \, \varsigma \over 1 + R_\infty^{-1} \, \kappa_0 \, \varsigma} .
\label{F2}
\end{eqnarray}
Equation~(\ref{F2}) yields the expression for $\varsigma$ as
\begin{eqnarray}
\varsigma = {{\rm Ri}_{\rm f} \over \kappa_0 \, (1 -  {\rm Ri}_{\rm f} / R_\infty)} .
\label{FF2}
\end{eqnarray}
In this case, the vertical share of TKE $A_z(\varsigma) \equiv E_z / E_{\rm K}$ reads
\begin{eqnarray}
A_z  &=&  A_z^{(0)}
+ {1- A_z^{(0)} \over 1 -  (1 + C_\Phi)^{-1} \, [(\kappa_0 \, \varsigma)^{-1} + {\rm R}_\infty^{-1}]}
\nonumber\\
&& \qquad + {2 A_z^{(0)} \over 1 + {\rm R}_\infty \,(\kappa_0 \, \varsigma)^{-1} } ,
\label{F9}
\end{eqnarray}
and the vertical profile of the turbulent Prandtl number
${\rm Pr}_{_{T}}(\varsigma)$  is given by:
\begin{eqnarray}
{\rm Pr}_{_{\rm T}} = {\rm Pr}_{_{\rm T}}^{(0)} \,  \left[1 - { C_\theta \, C_{\rm p} \over A_z
\left[{\rm R}_\infty^{-1} +   (\kappa_0 \, \varsigma)^{-1} -  (1 + C_\Phi)\right]}
\right]^{-1} .
\nonumber\\
\label{F3}
\end{eqnarray}
Note that the gradient Richardson number ${\rm Ri}$ and the flux Richardson number ${\rm Ri}_{\rm f}$ are related
as ${\rm Ri}(\varsigma) = {\rm Ri}_{\rm f}(\varsigma) \, {\rm Pr}_{_{\rm T}}(\varsigma)$.

Equations~(\ref{F1})--(\ref{F3}) are in agreement with Monin-Obukhov-Nieuwstadt similarity theories \citep{MO54,N84}.
In the Monin-Obukhov similarity theory \citep{MO54}, the turbulent fluxes of momentum $\tau$, heat $F_z$, the Obukhov length scale $L$ and other scalars are approximated by their
surface values, while the similarity theory by Nieuwstadt  \citep{N84} is extended  to
the entire stably stratified boundary layer  employing local $z$-dependent values of the turbulent
fluxes $\tau(z)$ and $F_z(z)$, and the length $L(z)$ instead of their surface values.

Using Eqs.~(\ref{ML13}) and~(\ref{FF2}), we can relate $\varsigma$ and $\tilde Z$ for stably stratified turbulence as
\begin{eqnarray}
\varsigma = {\tilde Z \, \tilde E_{\rm K}^{1/2} \over \kappa_0 \, (1 + C_\Phi -  \tilde Z \, \tilde E_{\rm K}^{1/2} / R_\infty)} .
\label{MFF2}
\end{eqnarray}
For the surface layer  ($\tilde Z \ll 1$) of the stably stratified turbulence, the dimensionless height is $\varsigma = z/L$,
and the normalised TKE is $\tilde E_{\rm K} \approx 1$ [see Eq.~(\ref{A7})].
Therefore,  Eq.~(\ref{MFF2}) in this case is reduced to
\begin{eqnarray}
{z \over L} = {\tilde Z  \over \kappa_0 \, (1 + C_\Phi)} .
\label{MMFF2}
\end{eqnarray}
This equation coincides with Eq.~(\ref{LL10}) derived for the low part ($|\tilde Z| \ll 1$)
of the surface layer in convective turbulence.

\section{Surface layers in convective turbulence}
\label{sec:5}

\begin{figure}
\centering
\includegraphics[width=9.5cm]{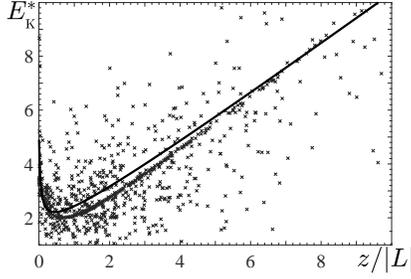}
\caption{\label{Fig2} The normalized turbulent kinetic energy $E_{\rm K}^{\ast} = E_{\rm K}/u_\ast^2$
versus $z / L$ obtained in the EFB theory (solid line) which is compared with the data
obtained from meteorological observations at the Eureka station located in the Canadian territory of Nunavut \citep{GP18}
in the conditions of long-lived convective boundary layer typical of the arctic summer.
}
\end{figure}

\begin{figure}
\centering
\includegraphics[width=8.0cm]{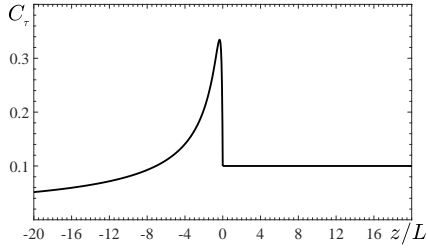}
\caption{\label{Fig3} The function $C_\tau$
versus $z / L$ for convective and stably stratified turbulence.
}
\end{figure}

In this section we apply results obtained in Section~\ref{sec:3} to
convective turbulence. In this case, the nonlinear equation for the vertical profile
of the normalized TKE, $\tilde E_{\rm K}(\tilde Z) = E_{\rm K}(\tilde Z)/E_{\rm K0}$
is given by Eq.~(\ref{MA6}).
In  Fig.~\ref{Fig2} we show the normalized turbulent kinetic energy $E_{\rm K}^{\ast} = E_{\rm K}/u_\ast^2$
versus $z / L$ obtained in the EFB theory. This dependence has been  compared
with the data \citep{ZRK21} obtained from meteorological observations at the Eureka station located in the Canadian territory of Nunavut \citep{GP18}
in the conditions of long-lived convective boundary layer typical of the arctic summer.
The better agreement between theoretical predictions and the observation data is achieved when $C_\tau$ is the following
function of $z / L$ (see  Fig.~\ref{Fig3}):
\begin{eqnarray}
C_\tau = \left(0.1 + {1.72 \, |X(z)| \over 1 + 2 |X(z)|} \right) \, \left(1 + {4 z \over |L|}\right)^{-2/3} ,
\label{L1}
\end{eqnarray}
where $X(z)= {\rm Ri}_{\rm f}(z)/ {\rm R}_\infty$. We remind that $C_\tau$ is related to
the effective dissipation time scale of the Reynolds stress.

Asymptotic solutions of  Eq.~(\ref{MA6}) for the normalized TKE, $\tilde E_{\rm K}(\tilde Z)$ are given
by  Eq.~(\ref{A7}) for a lower part ($ |\tilde Z|  \ll 1$) of the surface convective layer, and by
Eq.~(\ref{A8}) for an upper part ($ |\tilde Z|  \gg 1$) of the surface convective layer.
Below we present asymptotic formulas for various turbulent characteristics based on
Eqs.~(\ref{TT10}), (\ref{ML13})--(\ref{FFF8}), (\ref{MDD5})--(\ref{FFA12}), (\ref{MCC8})
and (\ref{MCRC11})--(\ref{R11}) for the lower and upper parts of the surface layer in convective turbulence.
In particular, the turbulence characteristics for a lower part ($ |\tilde Z|  \ll 1$) of the surface convective layer are given by
\begin{itemize}
\item{
the flux Richardson number,
\begin{eqnarray}
{\rm Ri}_{\rm f} = -  (1 + C_\Phi)^{-1} \,  \left|\tilde Z\right| ,
\label{ML13a}
\end{eqnarray}
}
\item{
the large-scale shear,
\begin{eqnarray}
S  ={u_\ast  \over \kappa_0 \, z}  ,
 \label{A11a}
\end{eqnarray}
}
\item{
the turbulent viscosity,
\begin{eqnarray}
K_{\rm M} = \kappa_0 \, u_\ast \, z  ,
 \label{A10a}
\end{eqnarray}
}
\item{
the vertical share of TKE,
\begin{eqnarray}
A_z({\rm Ri}_{\rm f})  =  A_z^{(0)}  + \Big(1- A_z^{(0)} \Big)  \,  \left|\tilde Z\right|   ,
\label{MCRC11a}
\end{eqnarray}
}
\item{
the turbulent Prandtl number,
\begin{eqnarray}
{\rm Pr}_{_{\rm T}} = {\rm Pr}_{_{\rm T}}^{(0)} \,  \left[1 - {C_\theta \, C_{\rm p} \over A_z^{(0)} \,  (1 + C_\Phi)} \,   \left|\tilde Z\right| \right],
\label{C14a}
\end{eqnarray}
}
\item{
 the level of temperature fluctuations,
\begin{eqnarray}
{E_\theta \over \theta_\ast^2} = C_{\rm p} \, \left(2 C_\tau \, A_z^{(0)}\right)^{-1/2}  \,   {\rm Pr}_{_{\rm T}}^{(0)}
\,  \left(1 - C_{\rm E} \,   \left|\tilde Z\right|  \right),
 \label{AA9a}
\end{eqnarray}
}
\item{
the vertical gradient of the mean potential temperature,
\begin{eqnarray}
\nabla_z \overline{\Theta} = -  {\theta_\ast  \, {\rm Pr}_{_{\rm T}}^{(0)} \over \kappa \, z} ,
 \label{FFA12a}
\end{eqnarray}
}
\item{
the ratio of the horizontal and vertical turbulent fluxes of potential temperature,
\begin{eqnarray}
{F_i \over F_z}= - C_{\rm F}  \, \left(1 + {\rm Pr}_{_{\rm T}}^{(0)}\right) \,  \left(2 \, C_\tau \, A_z^{(0)}\right)^{-1/2}  ,
\label{MCC8a}
\end{eqnarray}
}
\item{
the horizontal components of TKE,
\begin{eqnarray}
E_x=E_y  = {1 \over 2} \,  E_{\rm K0} \, \left(1 - A_z^{(0)}\right) \,  \left(1 - {1 \over 2} \,  \left|\tilde Z\right|  \right) ,
\label{R11a}
\end{eqnarray}
}
\end{itemize}
where
\begin{eqnarray}
C_{\rm E} = {1 \over 6} \left[1 + 2 \left(A_z^{(0)}\right)^{-1}\right] + {C_\theta \, C_{\rm p} \over A_z^{(0)} \,  (1 + C_\Phi)} .
\label{AAA9a}
\end{eqnarray}
In Eq.~(\ref{A11a}) we take into account that for the surface layer in convective turbulence,
the vertical integral turbulent scale, $\ell_z = C_\ell \, z$ and in Eq.~(\ref{MCC8a})  we consider the
case when the mean velocity $\meanU_i$ is directed along the $x$-axis.

For an upper part ($ |\tilde Z|  \gg 1$) of the surface convective layer, the turbulence characteristics are given by
\begin{itemize}
\item{
the flux Richardson number,
\begin{eqnarray}
{\rm Ri}_{\rm f} = -  (1 + C_\Phi)^{-1} \,  \tilde Z^{4/3} ,
\label{ML13b}
\end{eqnarray}
}
\item{
the large-scale shear,
\begin{eqnarray}
S  ={u_\ast  \over |L|}\,   (1 + C_\Phi) \, \tilde Z^{-4/3}  ,
 \label{A11b}
\end{eqnarray}
}
\item{
the turbulent viscosity,
\begin{eqnarray}
K_{\rm M} = (1 + C_\Phi)^{-1} \, u_\ast \, |L|  \, \tilde Z^{4/3}  ,
 \label{A10b}
\end{eqnarray}
}
\item{
the normalized vertical integral scale $\ell_z$,
\begin{eqnarray}
{\ell_z  \over |L| }=  (2 C_\tau)^{-3/4}    \, \tilde Z^{4/3}  ,
\label{FF8b}
\end{eqnarray}
}
\item{
the normalized TKE,
\begin{eqnarray}
{E_{\rm K} \over u_\ast^2} = (2 C_\tau)^{-1/2}  \, \tilde Z^{2/3}  ,
\label{C9b}
\end{eqnarray}
}
\item{
the vertical share of TKE,
\begin{eqnarray}
A_z({\rm Ri}_{\rm f})  &=&  1 - \Big(1- A_z^{(0)} \Big)  \,  \tilde Z^{-4/3} ,
\label{MCRC11b}
\end{eqnarray}
}
\item{
the turbulent Prandtl number,
\begin{eqnarray}
{\rm Pr}_{_{\rm T}} = {\rm Pr}_{_{\rm T}}^{(\infty)}\, \left[1 - \left(1-{{\rm Pr}_{_{\rm T}}^{(\infty)}
\over  {\rm Pr}_{_{\rm T}}^{(0)}} \right) \,  \tilde Z^{-4/3}\right],
\label{C14b}
\end{eqnarray}
}
\item{
 the level of temperature fluctuations $E_\theta/ \theta_\ast^2$,
\begin{eqnarray}
{E_\theta \over \theta_\ast^2} = C_{\rm p} \, \, (2 C_\tau)^{-1/2}   \,   {\rm Pr}_{_{\rm T}}^{(\infty)} \,  \tilde Z^{-2/3} ,
 \label{AA9b}
\end{eqnarray}
}
\item{
the vertical gradient of the mean potential temperature,
\begin{eqnarray}
\nabla_z \overline{\Theta}  = - {\theta_\ast \over |L|} \, {\rm Pr}_{_{\rm T}}^{(\infty)} \,  \tilde Z^{-4/3},
 \label{FFA12bb}
\end{eqnarray}
}
\item{
the ratio of the horizontal and vertical turbulent fluxes of potential temperature,
\begin{eqnarray}
{F_x \over F_z}= - C_{\rm F}  \, \left(1 + {\rm Pr}_{_{\rm T}}\right) \,  \left(2 \, C_\tau  \right)^{-1/2} \,   \tilde Z^{-2/3} ,
\label{MCC8b}
\end{eqnarray}
}
\item{
the horizontal components of TKE,
\begin{eqnarray}
E_x=E_y  = {1 \over 2} \, E_{\rm K0} \, \left(1 - A_z^{(0)}\right) \,  \tilde Z^{-2/3} ,
\label{R11b}
\end{eqnarray}
}
\end{itemize}
where
\begin{eqnarray}
{\rm Pr}_{_{\rm T}}^{(\infty)} = {\rm Pr}_{_{\rm T}}^{(0)} \, \left[1+ {C_\theta \, C_{\rm p} \over 1 + C_\Phi}\right]^{-1},
\label{C14bb}
\end{eqnarray}
and in Eq.~(\ref{MCC8b})  we consider the case when the mean velocity $\meanU_i$ is directed
along the $x$-axis.

\begin{figure}
\centering
\includegraphics[width=8.0cm]{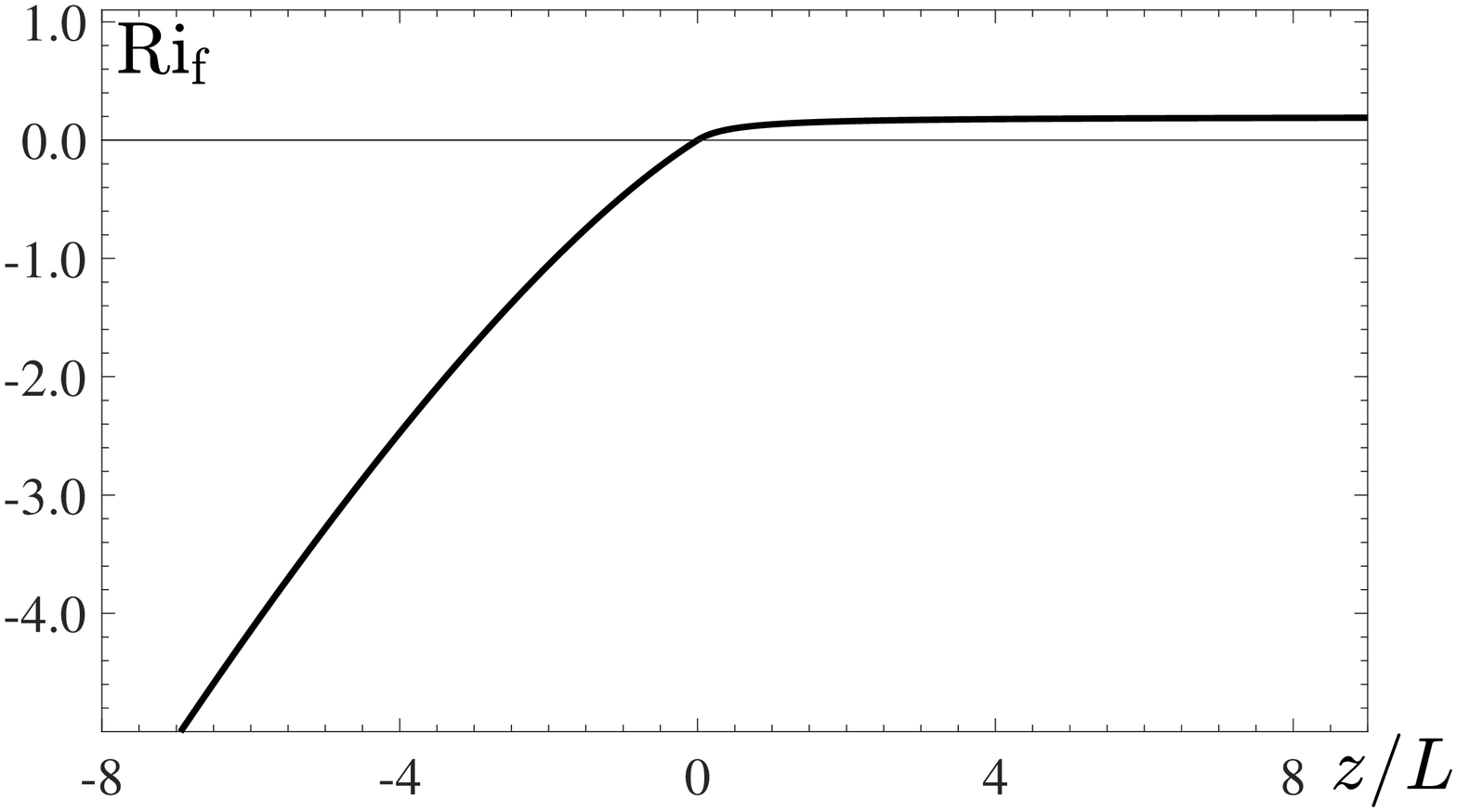}
\caption{\label{Fig4} The flux Richardson number ${\rm Ri}_{\rm f}$
versus $z / L$ for convective and stably stratified turbulence.
}
\end{figure}

\begin{figure}
\centering
\includegraphics[width=8.0cm]{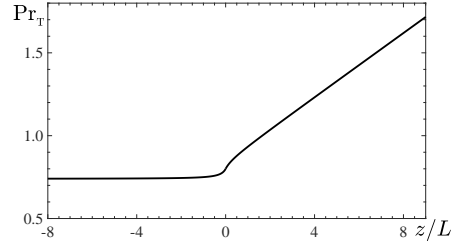}
\caption{\label{Fig5} The turbulent Prandtl number ${\rm Pr}_{_{T}}$
versus $z / L$ for convective and stably stratified turbulence.
}
\end{figure}

Substituting Eq.~(\ref{T10}) and the relation $\ell_z = C_\ell \, z$ into Eq.~(\ref{C9b}), we arrive at
the famous expression for the convective  turbulent energy,
\begin{eqnarray}
E_{\rm K} = C_{\rm c} \,  (\beta \, F_z \, z)^{2/3} ,
 \label{FFA12b}
\end{eqnarray}
obtained using dimension analysis in Ref.~ \cite{P32}.
Scalings for convective turbulence similar to Eqs.~(\ref{ML13b})--(\ref{A10b}) and~(\ref{AA9b})--(\ref{FFA12bb}) (where $\ell_z = C_\ell \, z$), were obtained
in Refs.~\cite{P32,O46,O71} using dimensional analysis (see also for a review books by \cite{MY71,MY75,Z91}).
The scalings similar to Eqs.~(\ref{MCC8b})--(\ref{R11b}) were derived using dimensional analysis in Refs.~\cite{Z71,Z73}.
Most of the above scalings are in agreement with the data of the atmospheric observations discussed in Ref.~\cite{KY91}.

\begin{figure}
\centering
\includegraphics[width=8.0cm]{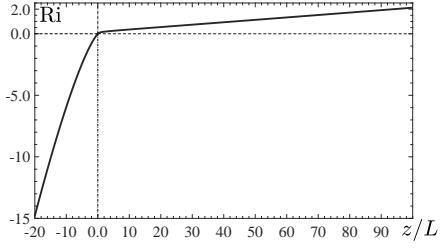}
\caption{\label{Fig6} The gradient Richardson number ${\rm Ri}$
versus $z / L$ for convective and stably stratified turbulence.
}
\end{figure}

\begin{figure}
\centering
\includegraphics[width=8.0cm]{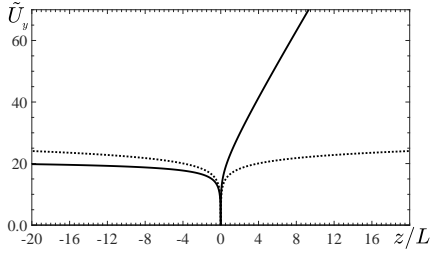}
\caption{\label{Fig7}
The normalised mean velocity $\tilde U_y = \meanU_y/u_\ast$  (solid line) versus $z / L$ for convective and stably stratified turbulence,
where the normalised height of the roughness elements is $z_\ast / L = 5.28 \times 10^{-4}$.
The dotted line corresponds to $\tilde U_y = \kappa^{-1} \, \ln (z / z_\ast)$.
}
\end{figure}

\begin{figure}
\centering
\includegraphics[width=8.0cm]{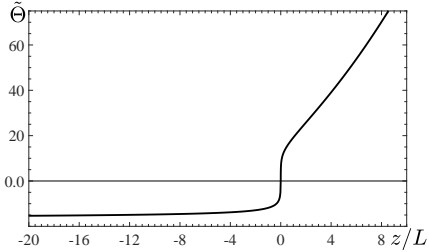}
\caption{\label{Fig8}
The normalised mean temperature difference $\tilde \Theta = (\meanT- \meanT_b)/\theta_\ast$
versus $z / L$ for convective and stably stratified turbulence,
where $\meanT_b$ is the mean temperature at the lower boundary.
}
\end{figure}

\begin{figure}
\centering
\includegraphics[width=8.0cm]{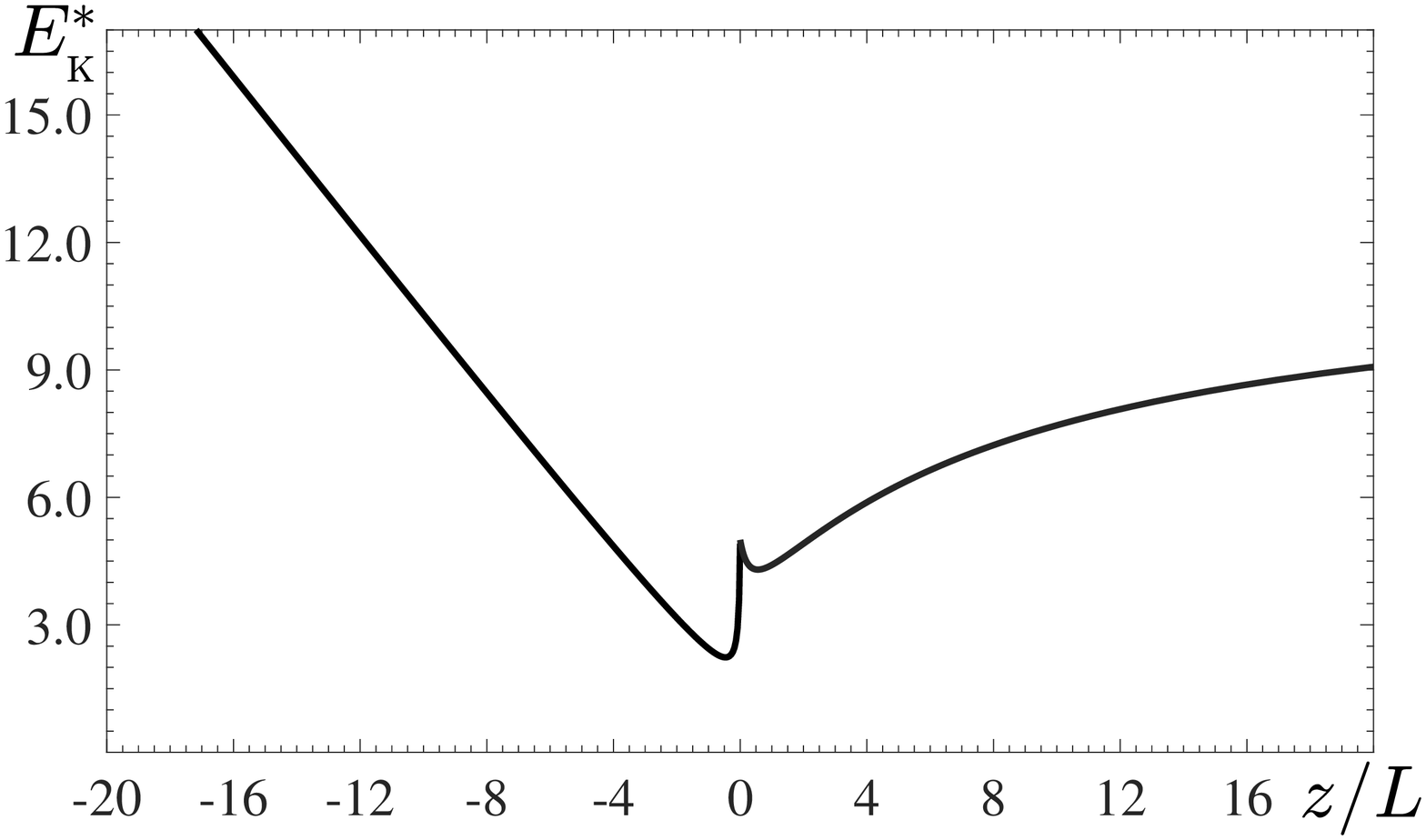}
\caption{\label{Fig9} The normalized turbulent kinetic energy $E_{\rm K}^{\ast} = E_{\rm K}/u_\ast^2$
versus $z / L$ for convective and stably stratified turbulence.
}
\end{figure}

\begin{figure}
\centering
\includegraphics[width=8.0cm]{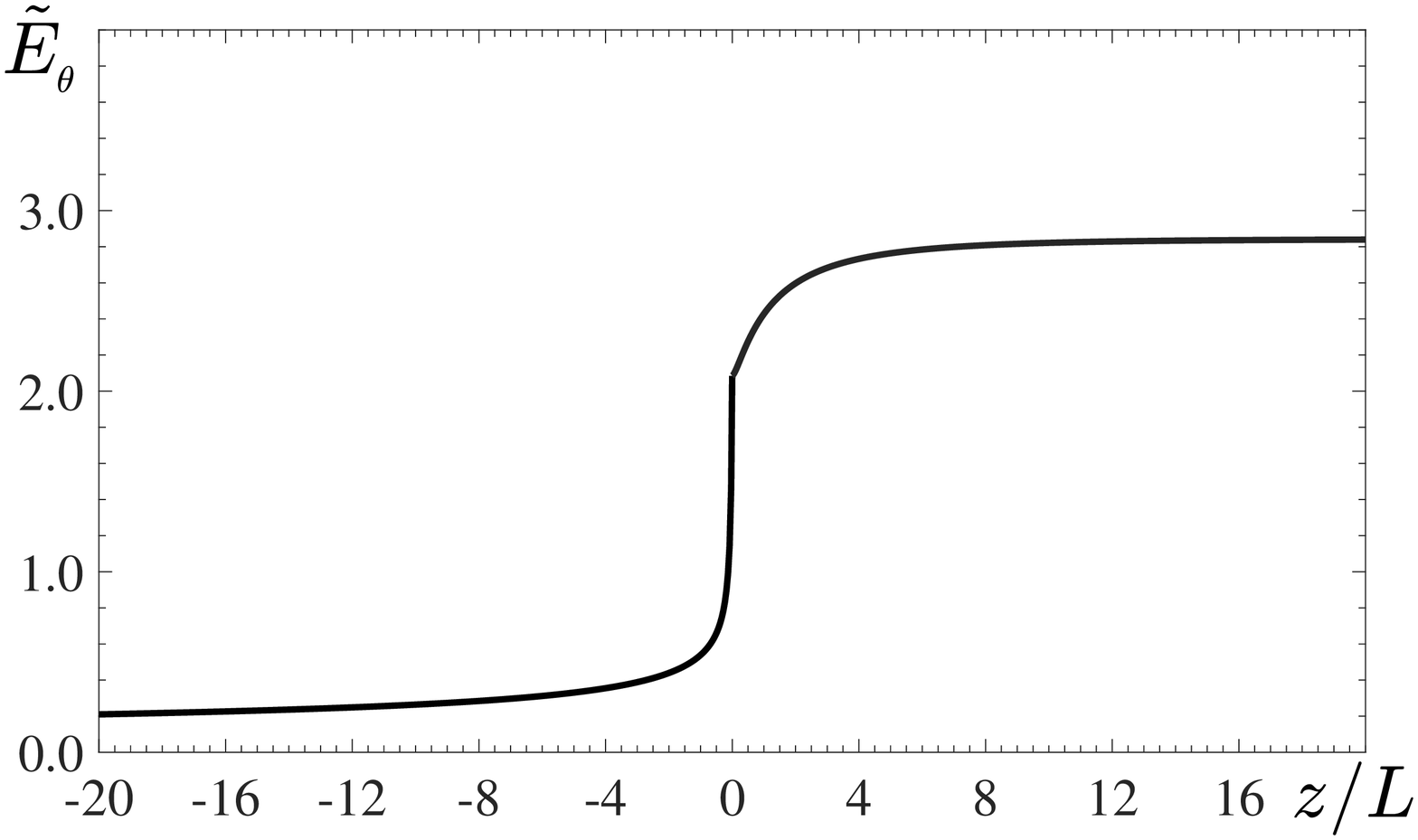}
\caption{\label{Fig10} The normalized intensity of potential temperature fluctuations $\tilde E_\theta=E_\theta / \theta_\ast^2$
versus $z / L$  for convective and stably stratified turbulence.
}
\end{figure}

\begin{figure}
\centering
\includegraphics[width=8.0cm]{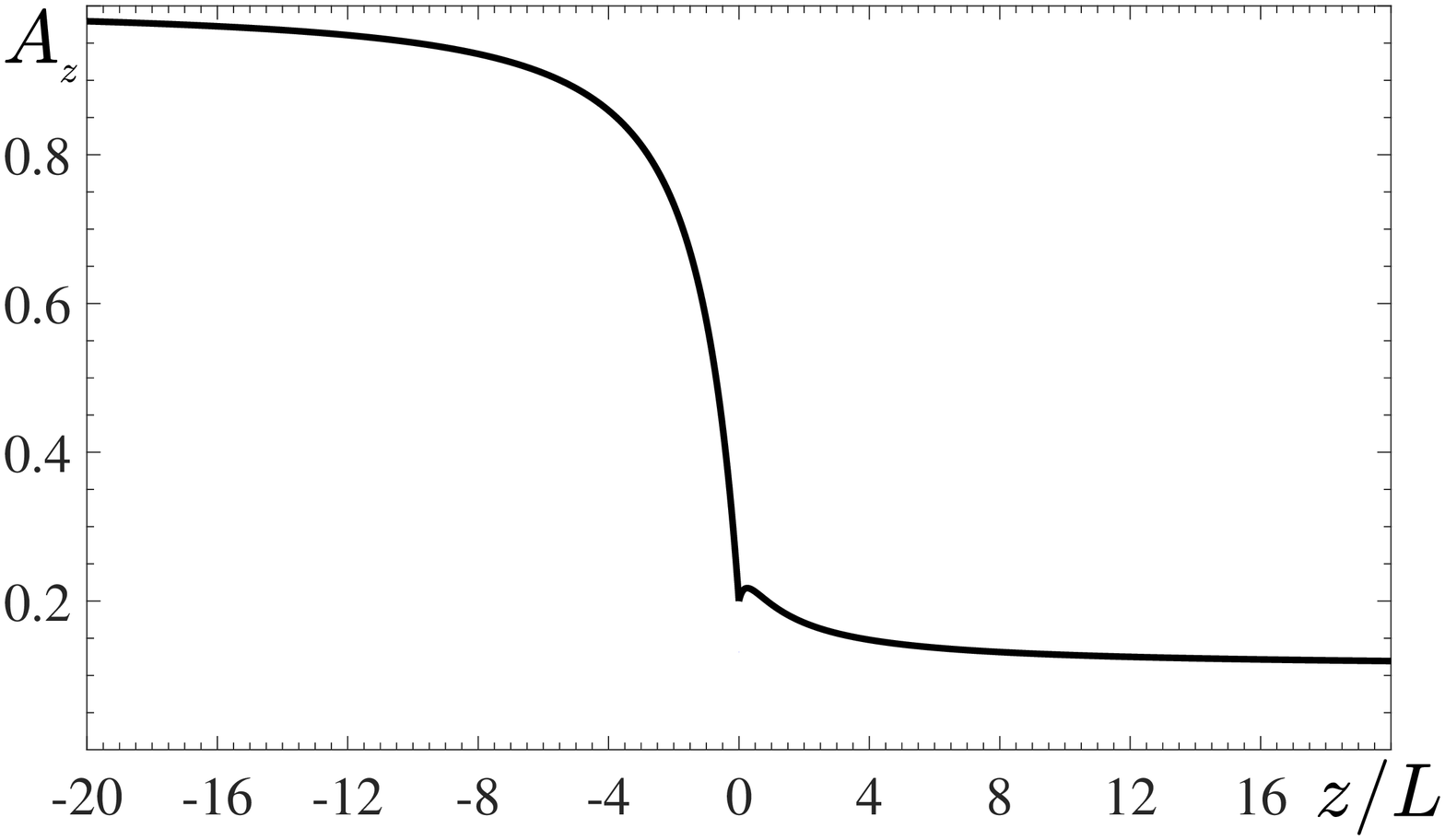}
\caption{\label{Fig11} The normalized vertical share  $A_z$ of turbulent kinetic energy
versus $z / L$ for convective and stably stratified turbulence.
}
\end{figure}

\begin{figure}
\centering
\includegraphics[width=8.0cm]{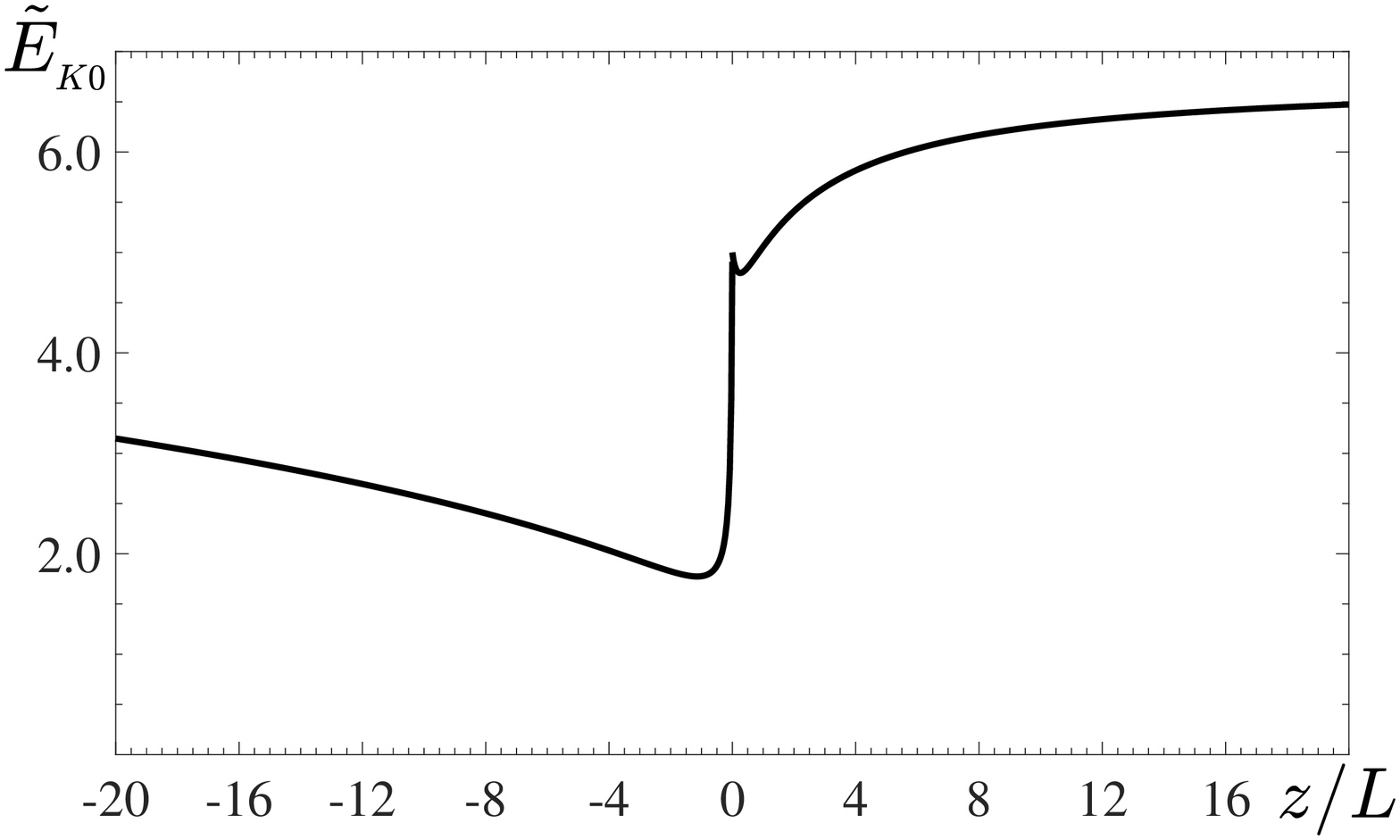}
\caption{\label{Fig12} The normalized turbulent kinetic energy $\tilde E_{\rm K0} = E_{\rm K0}/u_\ast^2$
versus $z / L$ for convective and stably stratified turbulence.
}
\end{figure}

\begin{figure}
\centering
\includegraphics[width=8.0cm]{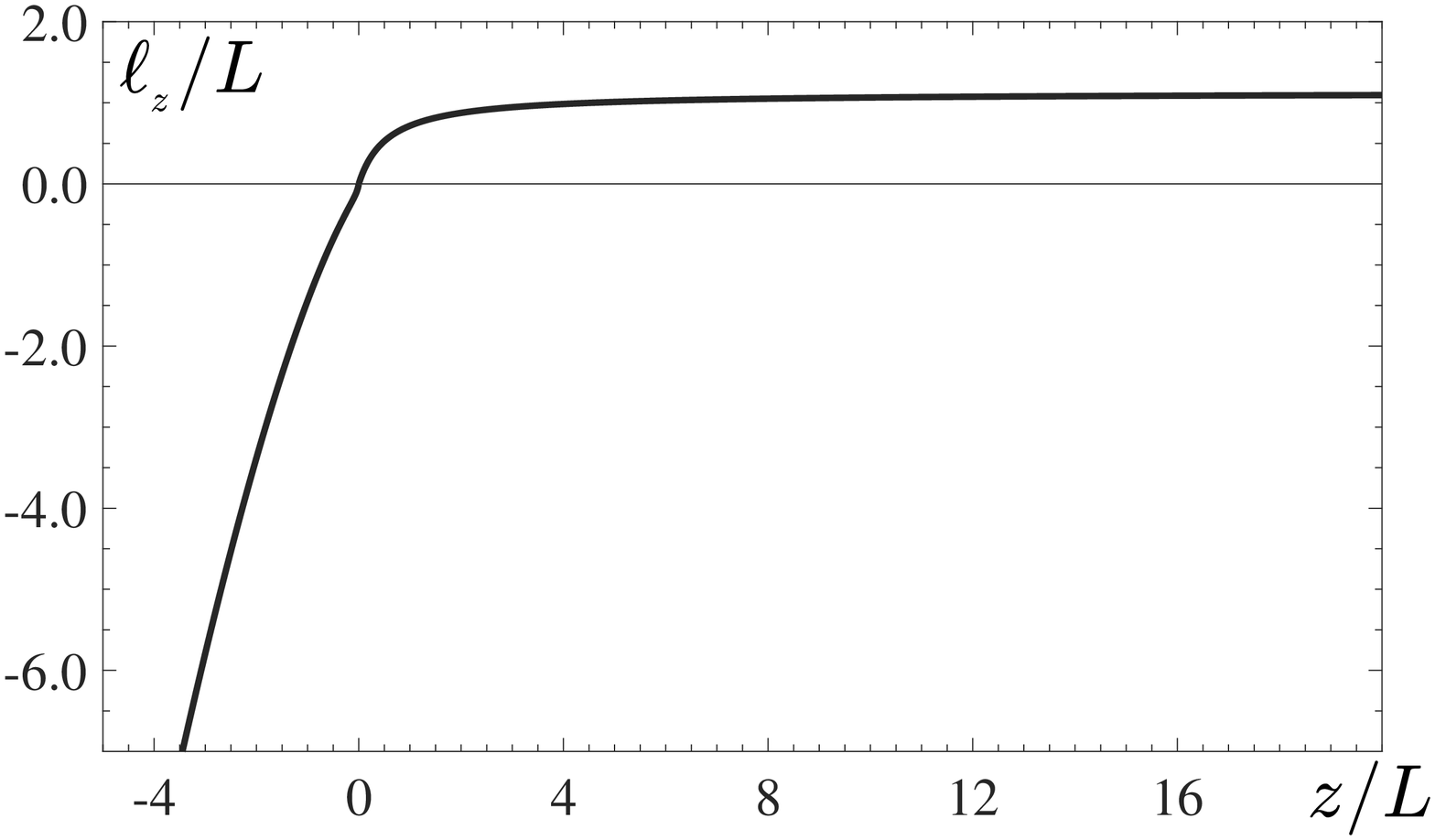}
\caption{\label{Fig13} The normalized vertical integral scale $\ell_z/L$
versus $z / L$ for convective and stably stratified turbulence.
}
\end{figure}

In Figs.~\ref{Fig4}--\ref{Fig13} we show vertical profiles of various turbulent characteristics for convective
($L<0$) and stably stratified ($L>0$) turbulence.
These dependencies are based on Eqs.~(\ref{A6}),  (\ref{MA6}), (\ref{LL10}), (\ref{TT10}), (\ref{ML13})--(\ref{FFF8}), (\ref{MDD5})--(\ref{FFA12}), (\ref{MCC8}),
(\ref{MMRC11})--(\ref{R11}) and~(\ref{MFF2}).
Three basic dimensional numbers,  ${\rm Ri}_{\rm f}$,
 ${\rm Pr}_{_{T}}$ and ${\rm Ri}$,  plotted in Figs.~\ref{Fig4}--\ref{Fig6},
 are related by the expression ${\rm Ri}={\rm Ri}_{\rm f} \, {\rm Pr}_{_{T}}$.

  Let us discuss the choice of the dimensionless empirical constants \citep{ZKR13,KRZ21}.
There are two well-known universal constants: the limiting value of the flux Richardson number
$R_\infty=0.2$ for an extremely strongly stratified turbulence (i.e., for ${\rm Ri} \to \infty$)
and the turbulent Prandtl number ${\rm Pr}_{_{T}}^{(0)}=0.8$ for a nonstratified turbulence (i.e., for ${\rm Ri} \to 0$)
\citep{EKR96b,CH02,FO06}.
The constant $C_{\rm p}$ describes the deviation of the dissipation timescale of $E_\theta=\langle \theta^2 \rangle/2$
from the dissipation timescale of TKE.
The constant $C_\theta$ is given by $C_\theta=A_z^{(\infty)}[(1 - {\rm R}_\infty (1+ C_\phi )]  / (C_{\rm p}{\rm R}_\infty)$ [see Eq.~(\ref{MDD5})].
We use here the following values of the non-dimensional empirical constants: $C_{\rm p} = 0.417$, $C_\theta = 0.744$,
$C_\phi = 0.899$,  $\kappa_0= 0.4$, ${\rm Pr}_{_{\rm T}}^{(0)}=0.8$, $A_z^{(0)}=0.2$, $A_z^{(\infty)}=0.1$.
These constants are determined from the data of numerous
meteorological observations, laboratory experiments, direct numerical simulations (DNS) and large eddy simulations (LES) \citep{SF01,OH01,SR10,ZKR07,ZKR13,LK16,KK78,BB97,RK04,SK00,CM12,MV05,PC02,BAN02,UC02}.
The parameter $C_\tau=0.1$ for stably stratified turbulence and $C_{\rm F}=C_\tau/{\rm Pr}_{_{T}}^{(0)}= 0.125$.
For convective turbulence, the parameters $C_\tau({\rm Ri}_{\rm f})$ and $C_{\rm F}({\rm Ri}_{\rm f})=C_\tau({\rm Ri}_{\rm f})/{\rm Pr}_{_{T}}^{(0)}$ are the functions of the flux Richardson number ${\rm Ri}_{\rm f}$ (see the beginning of Section~V).

 Absolute values of  the gradient Richardson number ${\rm Ri}$ in convective turbulence are much larger
 than in stably stratified turbulence. The reason is that the large-scale shear in convective turbulence
 is much smaller than in stably stratified turbulence (see Fig.~\ref{Fig10}).
This is because TKE in convective turbulence is much larger than in stably stratified turbulence (see Fig.~\ref{Fig12}), because
 in convection, both, the buoyancy and large-scale shear produce turbulence.
Contrary, in stably stratified turbulence, the large-scale shear produces TKE,
while the buoyancy decreases TKE and produces the temperature fluctuations.

On the other hand, the normalized intensity of potential temperature fluctuations $\tilde E_\theta=E_\theta / \theta_\ast^2$
(see Fig.~\ref{Fig10})  in convective turbulence is much weaker than in stably stratified turbulence.
The latter is caused by a weak  gradient of the mean potential temperature  in convective turbulence in comparison
with that of stably stratified turbulence (see Fig.~\ref{Fig8}).
The vertical share  $A_z$ of turbulent kinetic energy in stably stratified turbulence is changed stronger
than in the surface layers of convective turbulence (see Fig.~\ref{Fig11}). Indeed, turbulence tends to be two-dimensional one
for very large gradient Richardson number in stably stratified turbulence,  i.e., $A_z$ becomes very small.
Contrary, in convection the buoyancy is dominated in the energy production in the upper part of the surface layer,
resulting in a strong increase of the vertical TKE, i.e., the vertical share $A_z \to 1$.

Since the normalized turbulent kinetic energy $\tilde E_{\rm K0} = E_{\rm K0}/u_\ast^2$
is inversely proportional to the vertical share  $A_z$, it changes significantly in stably stratified turbulence
in comparison with convective turbulence  (see Fig.~\ref{Fig12}).
In Fig.~\ref{Fig13} we show the normalized vertical integral scale $\ell_z/L$
versus $z / L$ for convective and stably stratified turbulence. In stably stratified turbulence, the vertical integral scale
reaches the Obukhov length scale at high gradient Richarson numbers. Contrary, in convective turbulence
the ratio $\ell_z/|L|$ is strongly increases with height.

\section{Conclusions}
\label{sec:6}

We develop the energy and flux budget theory for the atmospheric surface layers
in turbulent convection and stably stratified turbulence.
This theory uses the budget equations for turbulent energies and fluxes.
In the framework of this theory we determine the vertical profiles for all turbulent characteristics
and for the mean velocity and mean potential temperature.
In particular, we find the vertical profiles of
turbulent kinetic energy, the intensity of  turbulent potential temperature fluctuations, the vertical turbulent fluxes
of momentum and buoyancy (proportional to potential temperature), the integral turbulence scale,
the turbulent anisotropy, the turbulent Prandtl number and the flux Richardson number.

Since the large-scale shear in convective turbulence
 is much smaller than in stably stratified turbulence, the
 absolute values of  the gradient Richardson number
 in convective turbulence are much larger
 than in stably stratified turbulence.
This is natural result, since turbulent kinetic energy (produced by both, the buoyancy
and large-scale shear) in convective turbulence is much stronger than
in stably stratified turbulence.
On the other hand, the large-scale shear produces turbulent kinetic energy in stably stratified turbulence,
and the buoyancy decreases TKE and produces the temperature fluctuations.
In convective turbulence, the gradient of the mean potential temperature is usually small  in comparison
with  stably stratified turbulence. Therefore, potential temperature fluctuations
are much smaller than in stably stratified turbulence.
The vertical integral scale in stably stratified turbulence can
only reach the Obukhov length scale at high gradient Richarson numbers. On the other hand,
the vertical integral scale in convective turbulence can be much larger than the
absolute value of the Obukhov length scale.

\begin{acknowledgements}
This paper is dedicated to Prof. Sergej Zilitinkevich (1936-2021) who initiated this work and discussed the obtained results.
This research was supported in part by the PAZY Foundation of the Israel Atomic Energy Commission (grant No. 122-2020),
and the Israel Ministry of Science and Technology (grant No. 3-16516).
\end{acknowledgements}

\bigskip

\section*{Appendix 1: The model for the inter-component energy exchange terms $Q_{\alpha\alpha}$
for anisotropic turbulence in the horizontal plane}

In this Appendix, we discuss the model for the inter-component energy exchange term $Q_{\alpha\alpha}$
for anisotropic turbulence in the horizontal plane.
In particular, the inter-component energy exchange terms $Q_{\alpha\alpha}$ are given by
\begin{eqnarray}
Q_{xx} = - {2(1 + C_{\rm r}) \over t_{\rm T}} \left[ E_x - {E_{\rm int} \over 3}  + A_z^{(0)}\left(C_1 + C_2 {{\rm Ri}_{\rm f} \over  R_\infty}\right)\right] ,
\nonumber\\
\label{AMAP1}
\end{eqnarray}
\begin{eqnarray}
Q_{yy} = - {2(1 + C_{\rm r}) \over t_{\rm T}} \left[ E_y - {E_{\rm int} \over 3}  - A_z^{(0)}\left(C_1 + C_2 {{\rm Ri}_{\rm f} \over  R_\infty}\right)\right] .
\nonumber\\
\label{AMAP2}
\end{eqnarray}
Using Eq.~(\ref{C4}) for $E_x$, we obtain that the horizontal shares $A_x = E_x/E_{\rm K}$
and $A_y = E_y/E_{\rm K}$ of TKE for stably stratified turbulence as
\begin{eqnarray}
A_x({\rm Ri}_{\rm f})  &=&  A_z^{(0)} \left[1 - C_1 - {{\rm Ri}_{\rm f} \over {\rm R}_\infty} \, (1 - C_2)\right]
\nonumber\\
&& + {1-3 A_z^{(0)}  \over 1 - {\rm Ri}_{\rm f}\,  (1 + C_\Phi)} ,
\label{AMRC11}
\end{eqnarray}
$A_y =1 - A_x - A_z$, and the vertical share $A_z$ is given by Eq.~(\ref{MMRC11}).
The free constants $C_1$ and $C_2$ are determined by the values $A_x^{(0)}$ at ${\rm Ri}_{\rm f}=0$ and $A_x^{(\infty)}$ at ${\rm Ri}_{\rm f} \to {\rm R}_\infty$.

\bigskip
\noindent
{\bf DATA AVAILABILITY}
\medskip

The data that support the findings of this study are available from the corresponding author
upon reasonable request.

\nocite{*}
\bibliography{Surface-layers-PF-references}

\end{document}